\documentclass[aps,prb,reprint,superscriptaddress]{revtex4-2}

\usepackage{graphicx}
\usepackage{dcolumn}
\usepackage{bm}
\usepackage{hyperref}
\usepackage{multirow}
\usepackage{amsmath}
\hypersetup{hypertex=true,
	colorlinks=true,
	linkcolor=blue,
	anchorcolor=blue,
	citecolor=blue}
\begin{document}
\title{Persistent Spin Dynamics in the Ising Triangular-lattice Antiferromagnet Ba$_6$Nd$_2$Ti$_4$O$_{17}$}

\author{C. Y. Jiang}
\affiliation{State Key Laboratory of Surface Physics, Department of Physics, Fudan University, Shanghai 200438, China}
\author{B. L. Chen}
\affiliation{State Key Laboratory of Surface Physics, Department of Physics, Fudan University, Shanghai 200438, China}
\author{K. W. Chen}
\affiliation{State Key Laboratory of Surface Physics, Department of Physics, Fudan University, Shanghai 200438, China}
\author{J. C. Jiao}
\affiliation{State Key Laboratory of Surface Physics, Department of Physics, Fudan University, Shanghai 200438, China}
\author{Y. Wang}
\affiliation{State Key Laboratory of Surface Physics, Department of Physics, Fudan University, Shanghai 200438, China}
\author{Q. Wu}
\affiliation{State Key Laboratory of Surface Physics, Department of Physics, Fudan University, Shanghai 200438, China}
\author{N. Y. Zhang}
\affiliation{State Key Laboratory of Surface Physics, Department of Physics, Fudan University, Shanghai 200438, China}
\author{M. Y. Zou}
\affiliation{State Key Laboratory of Surface Physics, Department of Physics, Fudan University, Shanghai 200438, China}
\author{P.-C. Ho}
\affiliation{Department of Physics, California State University, Fresno, California 93740, USA}
\author{O. O. Bernal}
\affiliation{Department of Physics and Astronomy, California State University, Los Angeles, California 90032, USA}
\author{L. Shu}
\email{leishu@fudan.edu.cn}
\affiliation{State Key Laboratory of Surface Physics, Department of Physics, Fudan University, Shanghai 200438, China}
\affiliation{Shanghai Research Center for Quantum Sciences, Shanghai 201315, China}


\date{\today}

\begin{abstract}
We report results of magnetic susceptibility, specific heat, and muon spin relaxation ($\mu$SR) measurements on the polycrystalline Ba$_6$Nd$_2$Ti$_4$O$_{17}$, a disorder-free triangular-lattice antiferromagnet. The absence of long-range magnetic order or spin freezing is confirmed down to 30~mK, much less than the Curie-Weiss temperature -1.8~K. The magnetic and specific heat measurements reveal the effective-1/2 spins are Ising-like. The persistent spin dynamics is determined down to 37~mK. Our study present a remarkable example of Ising spins on the triangular lattice, which remains magnetically disordered at low temperatures and potentially hosts a quantum spin liquid (QSL) ground state.

\end{abstract}

\maketitle

\section{\label{intro}INTRODUCTION}
Quantum spin liquid (QSL) is an exotic state in which spins are highly entangled and remain disordered even at zero temperature due to strong quantum fluctuation~\cite{balents2010QSL,anderson1987RVB,wen2002QSL,Takagi2019KQSL}. Beyond the traditional Landau's symmetry breaking paradigm, QSL is characterized by fractionalized spin excitations instead of an order parameter~\cite{Wen1989frac,Wen1991frac}. The triangular-lattice antiferromagnet with isotropic Heisenberg interactions was proposed to host a QSL ground state by Anderson in the resonating valence bond picture~\cite{Anderson1973QSL}, but a magnetically-ordered ground state was demonstrated subsequently even for a spin-1/2 system where the quantum effects are most significant~\cite{Huse1988Hesienberg}. However, such magnetic order is fragile, it can be melted by other interactions, such as next-nearest-neighbor couplings~\cite{Iqbal2016J1J2,Zhu2015J1J2}, spatially anisotropic exchange interactions~\cite{Luo2017anisotropy}, or magnetic anisotropy~\cite{Motrunich2005anisotropy,Yamamoto2014anisotropy}. These effects provide an alternative way to realize QSL.\par

Rare-earth (RE)-based frustrated lattices represent a habitat for QSL, since the strong atomic spin-orbit coupling (SOC) gives rise to both spatial and spin anisotropies~\cite{Li2016Kramerstheory,William2013SOC}. YbMgGaO$_4$ and chalcogenides family NaYb$Ch_2$ ($Ch$ = O, S, Se) have attracted much attention due to the potential QSL properties~\cite{li2015YMGO,shen2016YMGO,li2017YMGO,Ding2019NaYbO2,Baenitz2018NaYbS2,Zhu2023NaYbSe2}. SOC together with the crystal electric field (CEF) leads to a Kramers ground-state doublet characterized by an effective 1/2 spin in these magnets~\cite{Li2016Kramerstheory}. However, the effective 1/2 spin basically hosts easy-plane magnetic anisotropy for the RE-based triangular lattices that have been extensively studied so far.\par

Recently, a Nd-based triangular lattice NdTa$_7$O$_{19}$ has been reported to be a potential QSL candidate~\cite{Arh2022NdTa7O19}. The effective spin-1/2 Nd$^{3+}$ moments are Ising-like, giving rise to spin excitations down to 40~mK in NdTa$_7$O$_{19}$. As a triangular lattice with Ising spins, NdTa$_7$O$_{19}$ provides a promising way to realize QSL and other novel quantum states, which has not been studied before due to the limitation of materials availability~\cite{Khatua2023review}. Actually, after early work of classical Ising model on the triangular lattice which led to a classical spin liquid state was proposed by Wannier~\cite{Wannier1950triangular}, many theoretical studies have predicted exotic states for Ising spins on the triangular lattice by introducing quantum fluctuations~\cite{Fazekas1974triangular,Moessner2001triangular}. Synthesizing and studying new Ising triangular-lattice materials is urgently needed.\par 

Here, we report comprehensive magnetic susceptibility, specific heat, and $\mu$SR studies on the polycrystalline Ba$_6$Nd$_2$Ti$_4$O$_{17}$, which is a disorder-free Nd$^{3+}$ triangular lattice. That the spin anisotropy is Ising-like is supported by magnetization and specific heat results. The absence of any magnetic order is confirmed down to 37~mK by ZF-$\mu$SR measurements despite an Curie-Weiss temperature of -1.8~K. A two-level Schottky anomaly is observed in the magnetic specific heat with applied magnetic fields, indicating the spin is effectively 1/2 at low temperatures. The magnetically disordered state with effective-1/2 spin is further studied by LF-$\mu$SR measurements, which reveal persistent spin dynamics down to 30~mK. Our results demonstrate that Ba$_6$Nd$_2$Ti$_4$O$_{17}$ is a potential QSL candidate.\par

\section{\label{expdetail}EXPERIMENTAL DETAILS}
Polycrystalline Ba$_6$Nd$_2$Ti$_4$O$_{17}$ was synthesized by the solid state reaction~\cite{Kuang2002growth}. The starting materials were BaCO$_3$, TiO$_2$, Nd$_2$O$_3$, they were first dried overnight at 700~$^{\circ}$C. Then stoichiometric amounts of reagents were mixed, thoroughly grounded, and pre-reacted at 900~$^{\circ}$C for 12 hours. The samples were finally obtained after reground and heated at 1250~$^{\circ}$C for 48 hours in air with repeating 2-3 times. The powder X-ray diffraction (XRD) data were collected using a Bruker D8 advanced X-ray diffraction spectrometer ($\lambda$ = 1.5418~$\rm{\AA}$) at room temperature. The structural refinements were performed using software package \textsc{FULLPROF} suite~\cite{Fullprof}.\par
Magnetization and direct-current (dc) magnetic susceptibility  were measured using a Magnetic Property Measurement System (MPMS, Quantum Design). The specific heat measurements were performed by the adiabatic relaxation method in a Physical Property Measurement System (PPMS) (DynaCool, Quantum Design) equipped with Helium-3 option. The $\mu$SR measurements were carried out in the LAMPF spectrometer at the M20 beamline, and DR spectrometer at the M15 beamline, TRIUMF, Vancouver, Canada. The samples were mounted on a silver holder in the DR spectrometer and encased in thin silver tape in the LAMPF spectrometer. The $\mu$SR data were analyzed by using the \textsc{MUSRFIT} software package~\cite{MuSRfit}.
\section{\label{results}RESULTS}
\subsection{\label{str}Structure}
The polycrystalline Ba$_6$Nd$_2$Ti$_4$O$_{17}$ has been synthesized for some time, yet its physical magnetic properties have not been investigated extensively so far~\cite{Kuang2002growth}. The crystal structure of Ba$_6$Nd$_2$Ti$_4$O$_{17}$ (space group $P6_{3}/mmc$) is shown in Fig.~\ref{str}(a). Nd$^{3+}$ ions are coordinated
with six nearest neighboring O atoms, NdO$_6$ octahedra on the same layer are linked to each other by TiO$_4$ tetrahedra and TiO$_6$ octahedra. The Nd$^{3+}$ ions form triangular lattice in the $ab$ plane, as shown in Fig.~\ref{str}(b). The triangular-lattice Nd$^{3+}$ layers stacking along $c$ axis are separated by two layers of TiO$_4$ tetrahedra or Ti$_2$O$_9$ dimers and Ba-O polyhedrons, resulting in two different interlayer distances of Nd$^{3+}$ layers, $d_1$=7.40~$\rm{\AA}$, $d_2$=7.56~$\rm{\AA}$. TiO$_4$ tetrahedra between Nd$^{3+}$ layers with interlayer distance $d_2$ are not connected with each other. Thus, Nd$^{3+}$ ions with interlayer distance $d_2$ lack the path of interlayer superexchange interaction, the interlayer interaction are negligible. While the Nd$^{3+}$ ions with interlayer distance $d_1$ are linked by Ti$_2$O$_9$ dimers, the path of interlayer superexchange interaction is Nd-O-Ti-O-Ti-O-Nd. The Nd$^{3+}$ ions within the triangular plane are linked by TiO$_4$ tetrahedra, and the path of intralayer interaction is Nd-O-Ti-O-Nd. Therefore, the intralayer interaction is dominant in Ba$_6$Nd$_2$Ti$_4$O$_{17}$.\par
\begin{figure}[h]
	\centering
	\includegraphics[height=5.5cm,width=8.5cm]{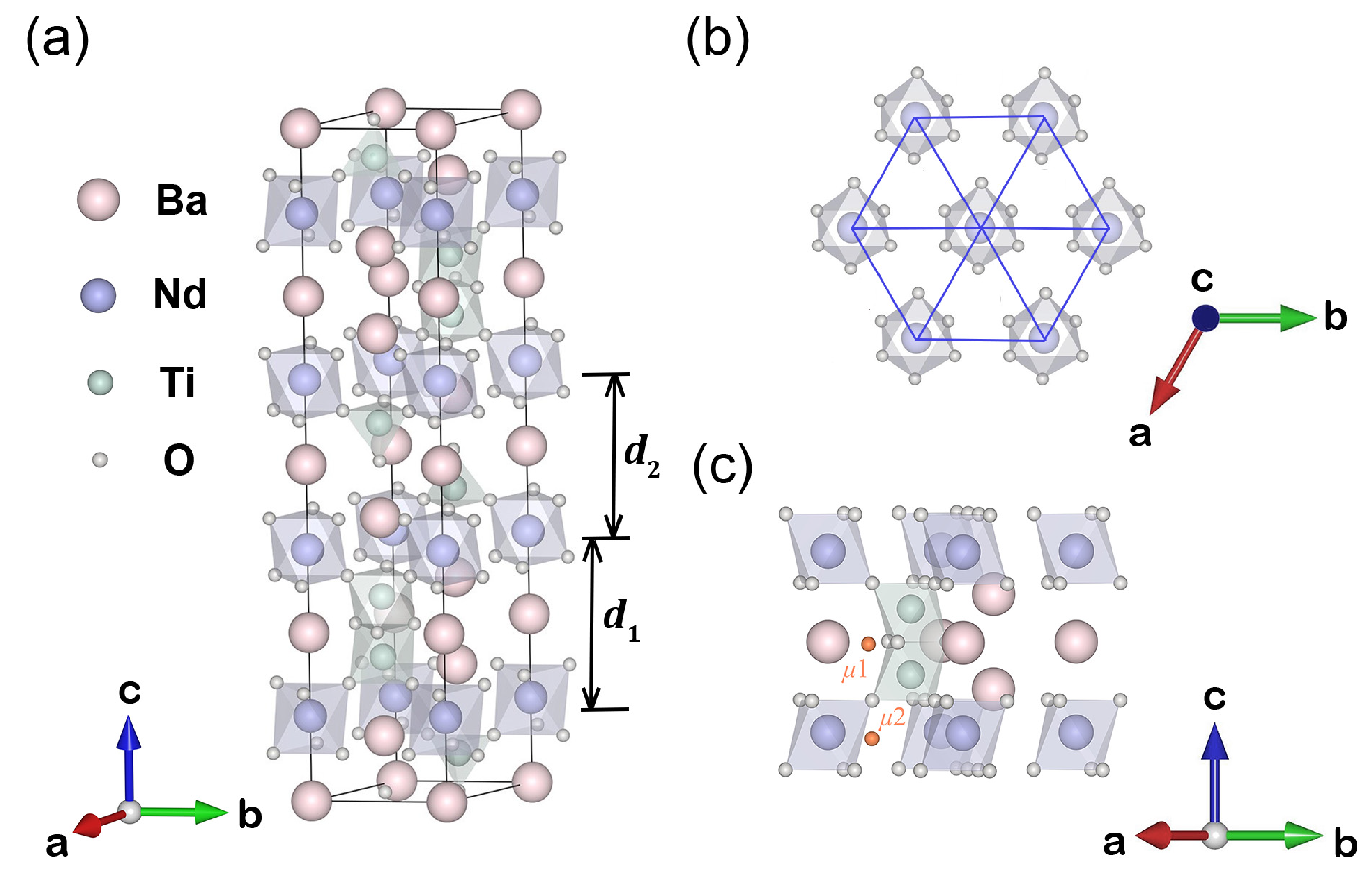}
	\caption{\label{str}(a) The crystal structure of Ba$_6$Nd$_2$Ti$_4$O$_{17}$ in one unit cell. (b) Top view of the triangular lattice formed by Nd$^{3+}$ ions. (c) The two possible muon stopping sites $\mu$1 and $\mu$2 are denoted by orange spheres. $\mu$1 is close to the Nd$^{3+}$ triangular plane, whereas $\mu$2 is located between two Nd$^{3+}$ triangular layers.}
\end{figure}

\begin{figure}[h]
	\centering
	\includegraphics[height=5cm,width=8cm]{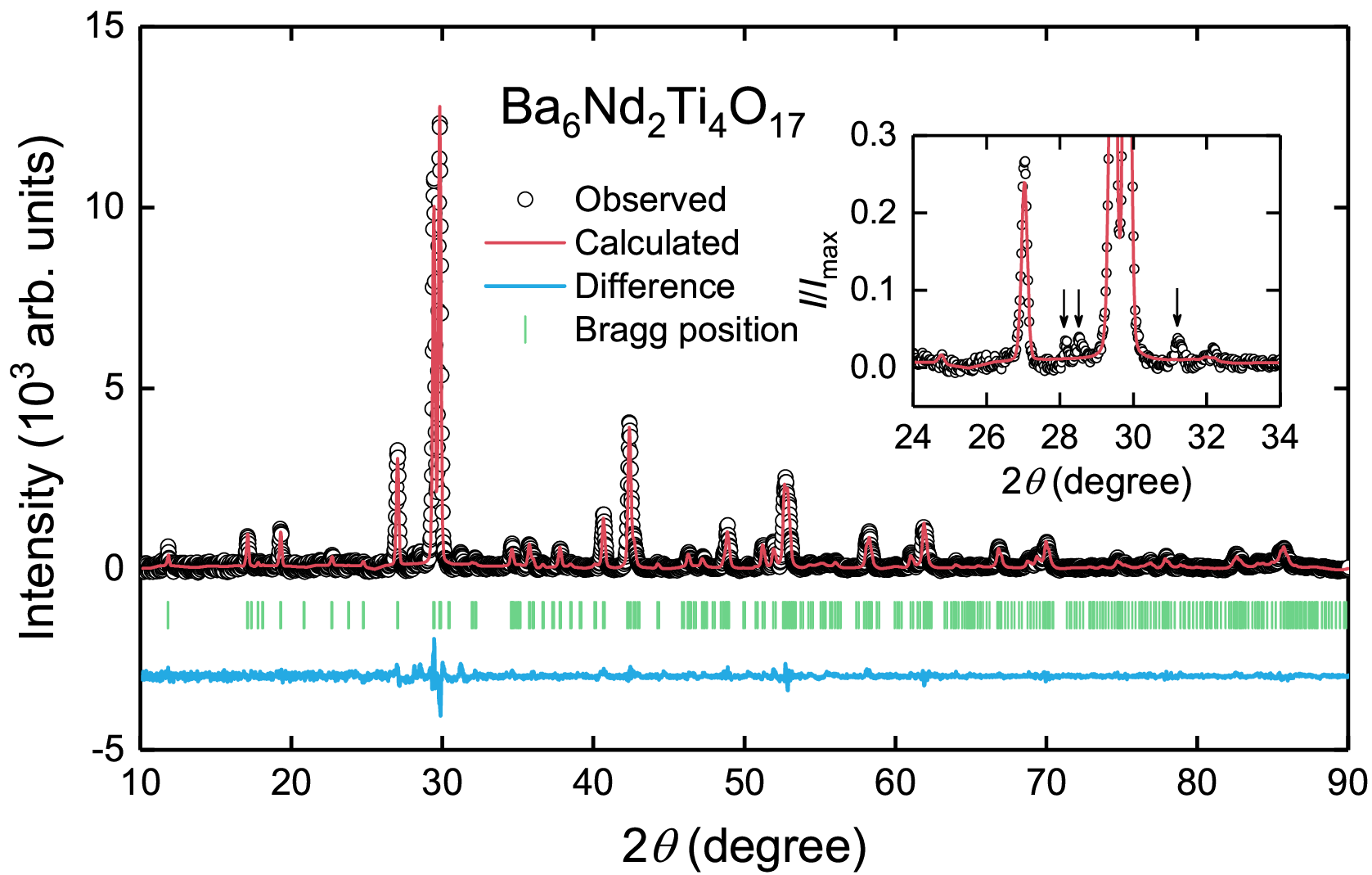}
	\caption{\label{xrd}The XRD pattern and Rietveld refinement results of Ba$_6$Nd$_2$Ti$_4$O$_{17}$. The inset shows an enlarged view of the region $2\theta=24^{\circ}$-$34^{\circ}$. The arrows sign the tiny peaks caused by unknown impurities.}
\end{figure}

The powder XRD measurements were performed to determine the sample quality. The XRD pattern and Rietveld refinements of polycrystalline Ba$_6$Nd$_2$Ti$_4$O$_{17}$ are shown in Fig.~\ref{xrd}, and the detailed refinement parameters are listed in Table.~\ref{RR}. The refinement discloses three tiny unexpected peaks at $2\theta=24^{\circ}$-$34^{\circ}$ as shown in the inset of Fig.~\ref{xrd}. However, none of the residual starting materials are matched with the tiny peaks, and Nd$^{3+}$/Ba$^{2+}$ or Nd$^{3+}$/Ti$^{4+}$ site-mixing disorder is also not matched. The absence of site-mixing disorder was also reported for single crystal Ba$_6$Nd$_2$Ti$_4$O$_{17}$~\cite{Song2024BTNO}. These unexpected peaks may originate from very few unknown intermediate impurities.
\begin{table}
	\caption{\label{RR}Rietveld refinement results for Ba$_6$Nd$_2$Ti$_4$O$_{17}$ XRD data measured at room temperature. The cell parameters are $\alpha=\beta=90^\circ$, $\gamma=120^\circ$, $a=b=5.992~\rm{\AA}$, $c=29.927~\rm{\AA}$. Overall B=0.33(6) ${\rm{\AA}}^{2}$. The space group is $P6_{3}/mmc$. $R_{wp}=4.00\%$, $R_{p}=2.91\%$, ${\chi}^{2}=2.40$.}
	\begin{ruledtabular}
		\begin{tabular}{ccccccc}
			Atom&Wyckoff position&$x$&$y$&$z$&Occ.\\
			\colrule
			Nd&4e&0&0&0.1263(2)&1\\
			Ba1&2a&0&0&0&1\\
			Ba2&4f&0.6667&0.3333&0.0861(3)&1\\
			Ba3&4f&0.3333&0.6667&0.1853(3)&1\\
			Ba4&2b&0&0&0.25&1\\
			Ti1&4f&0.3333&0.6667&0.0512(8)&1\\
			Ti2&4f&0.6667&0.3333&0.2079(6)&1\\
			O1&4f&0.3333&0.6667&-0.003(2)&1\\
			O2&12k&0.620(6)&0.810(3)&0.0755(7)&1\\
			O3&12k&0.349(7)&0.175(4)&0.1669(8)&1\\
			O4&6h&0.51(5)&0.019(8)&0.25&1\\
		\end{tabular}
	\end{ruledtabular}
\end{table}

\subsection{\label{mag}Magnetic properties}
\begin{figure}[h]
	\centering
	\includegraphics[height=9cm,width=6cm]{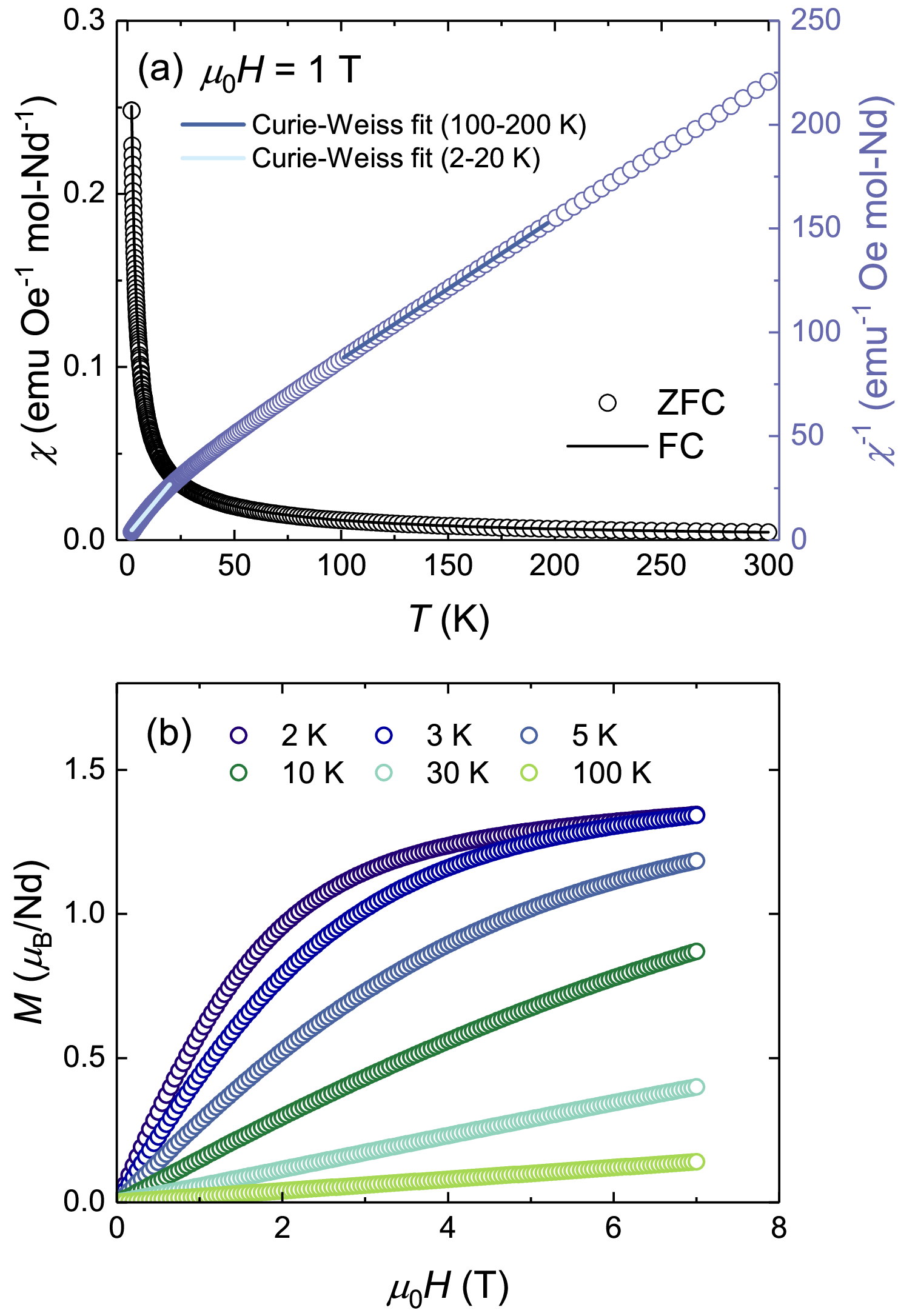}
	\caption{\label{chi}(a) The temperature dependence of dc magnetic susceptibility and its inverse of Ba$_6$Nd$_2$Ti$_4$O$_{17}$ at $\mu_0H$ = 1~T measured with ZFC and FC. The solid lines are Curie-Weiss fits in 2-10 K and 100-300 K respectively. (b) The isothermal magnetization at different temperatures down to 2~K.}
\end{figure}

The dc magnetic susceptibility $\chi$ and inverse susceptibility $\chi^{-1}$ of Ba$_6$Nd$_2$Ti$_4$O$_{17}$ are shown in Fig.~\ref{chi}(a). We measured the magnetic susceptibility with zero-field cooling (ZFC) and field cooling (FC) at $\mu_0H$ = 1~T. No sharp magnetic transition is observed down to 2~K and no obvious difference between ZFC and FC curves is detected, indicating the absence of spin ordering or freezing. The high-temperature Curie-Weiss fit (100-200~K) yields an effective magnetic moment $\mu_{\rm eff,H}$ = 3.43~$\mu_{\rm B}$, close to the theoretical value 3.62~$\mu_{\rm B}$ according to Hund's rule. The Curie-Weiss temperature yielded from high-temperature fitting is $\Theta_{\rm {CW},H}$ = -28.1(1)~K, suggesting the antiferromagnetic coupling of Nd$^{3+}$ moments. The magnetic susceptibility deviates from Curie-Weiss behavior as the temperature drops below 30~K. While at low-temperatures, Curie-Weiss law becomes valid again. The low-temperature Curie-Weiss fit (2- 20~K) yields a Curie-Weiss temperature $\Theta_{\rm {CW},L}$ = -1.8~K. The strength of exchange interaction between nearest Nd$^{3+}$ ions is roughly estimated $J_{\rm ex}\sim$1.2~K by the mean-field approximation using $J_{\rm ex}=3k_{B}\Theta_{\rm {CW},L}/(zJ_{\rm eff}(J_{\rm eff}+1))$, where $z=6$ is the number of nearest-neighbor Nd$^{3+}$ spins and $J_{\rm eff}=1/2$ is the effective spin~\cite{Greedan2001exchange}.
The effective magnetic moment of low-temperature Curie-Weiss fit is $\mu_{\rm eff,L}$ = 2.54~$\mu_{\rm B}$, much less than the theoretical value of free ion.\par
The isothermal magnetization up to 7~T at several temperatures is shown in Fig.~\ref{chi}(b). The magnetization at 2 K shows a non-linear field dependence above 1 T and tends to saturate above 4 T. The energy scale of 1 T magnetic field for Nd$^{3+}$ spins at low temperatures is $E=\mu_{\rm eff,L}B\sim$1.7~K, which is of the same order of the estimated exchange interaction strength $J_{\rm ex}\sim$1.2~K. The rough saturation value of magnetization is $\mu_{\rm sat}$ = 1.3~$\mu_{\rm B}$, about half the value of $\mu_{\rm eff,L}$ = 2.54~$\mu_{\rm B}$, which is a sign of powder-averaged Ising spins in polycrystalline sample~\cite{Bramwell2000Isingspin}. Our magnetization results are consistent with the previous results for polycrystalline Ba$_6$Nd$_2$Ti$_4$O$_{17}$~\cite{Song2024BTNO}. The field response of magnetization gets linear as the temperature increases up to 100~K as a consequence of strong thermal fluctuations.

\subsection{\label{hc}Specific heat}
\begin{figure}[h]
	\centering
	\includegraphics[height=14.5cm,width=6cm]{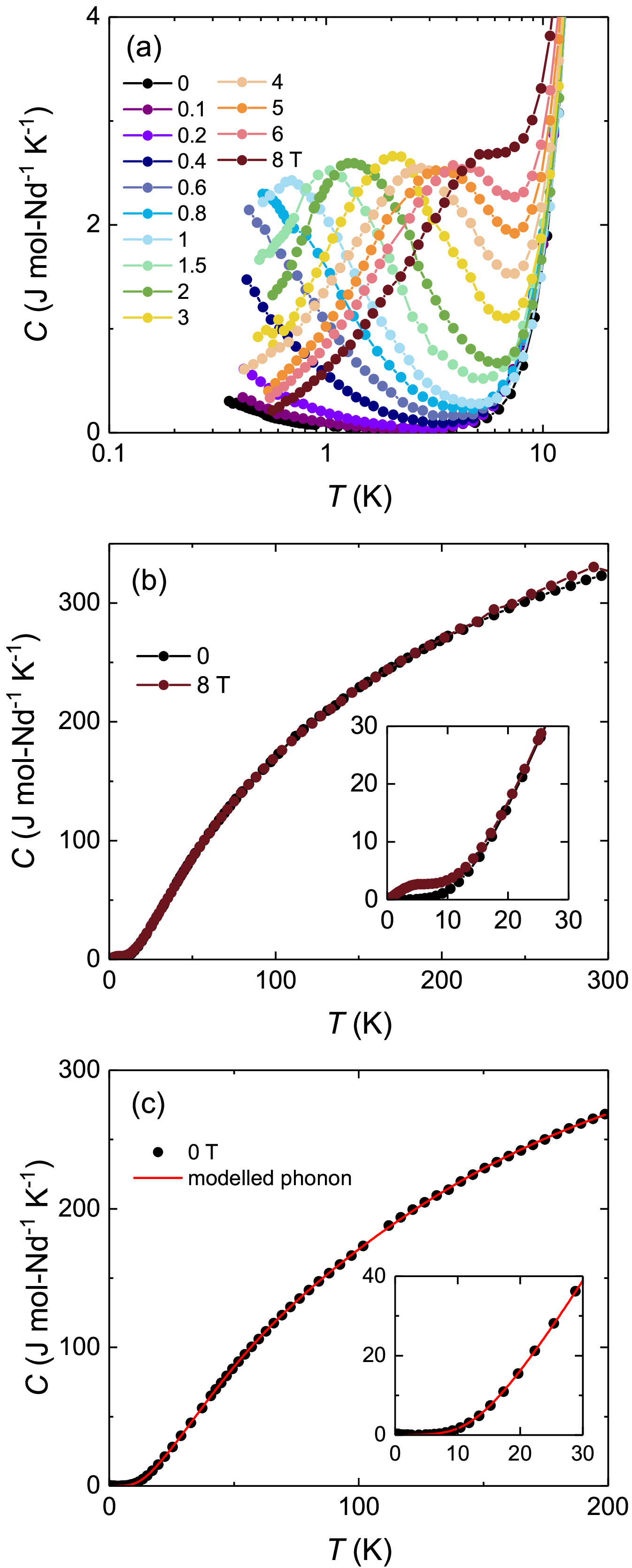}
	\caption{\label{specific1}(a) Low-temperature specific heat of polycrystalline Ba$_6$Nd$_2$Ti$_4$O$_{17}$ at different applied magnetic fields. (b) Specific heat measured at zero field and 8~T up to 300~K. The inset shows an enlarged view below 30~K. (c) The modelled phonon contribution using Eq.~\ref{Cph}. The inset shows detailed data below 30~K.}
\end{figure}

The lack of spin ordering or spin freezing is detected by magnetic susceptibility measurements down to 2~K. To further check the absence of magnetic phase transition and detect any magnetic excitations, we performed specific heat measurements on polycrystalline Ba$_6$Nd$_2$Ti$_4$O$_{17}$ down to 0.4~K. Fig.~\ref{specific1}(a) illustrates the specific heat under different magnetic fields at low temperatures. The absence of spin ordering is confirmed over the studied temperature range since not any sharp anomaly is observed. The specific heat up to 300~K at ZF and 8~T is shown in Fig.~\ref{specific1}(b). The data for ZF and 8~T basically overlap in the temperature range of 20~K$<T<$200~K, indicating the absence of excited CEF energy levels below 200~K. As seen in Fig.~\ref{specific1}(c), we obtain the phonon contribution by fitting ZF specific heat (2~-200~K) with a combination of two Debye and two Einstein functions
\begin{equation}
\begin{split}
	\label{Cph}
	C_{\rm ph}&=\sum_{i}^{2}f_{{\rm{D}}i}\left[9R(\frac{T}{\theta_{{\rm{D}}i}})^3\int_{0}^{\frac{\theta_{{\rm{D}}i}}{T}}\frac{x^4e^x}{(e^x-1)^2}dx\right]\\
	&+\sum_{i}^{2}f_{{\rm{E}}i}\left[3R(\frac{\theta_{{\rm{E}}i}}{T})^2\frac{e^{\frac{\theta_{{\rm{E}}i}}{T}}}{(e^{\frac{\theta_{{\rm{E}}i}}{T}}-1)^2}\right]
\end{split}
\end{equation}
The weight factors are fixed with the ratio $f_{\rm{D1}}\colon f_{\rm{D2}}\colon f_{\rm{E1}}\colon f_{\rm{E2}}=6\colon 15\colon 2\colon 6$, sum $\sum f_{{\rm{D}}i}+\sum f_{{\rm{E}}i}$ is equal to 29, which is the total number of atoms per formula unit of Ba$_6$Nd$_2$Ti$_4$O$_{17}$. The fitting yields Debye temperatures $\theta_{\rm{D1}}=172$~K, $\theta_{\rm{D2}}=726$~K, and Einstein temperatures $\theta_{\rm{E1}}=82$~K, $\theta_{\rm{E2}}=227$~K.\par
\begin{figure}[h]
	\centering
	\includegraphics[height=14cm,width=6cm]{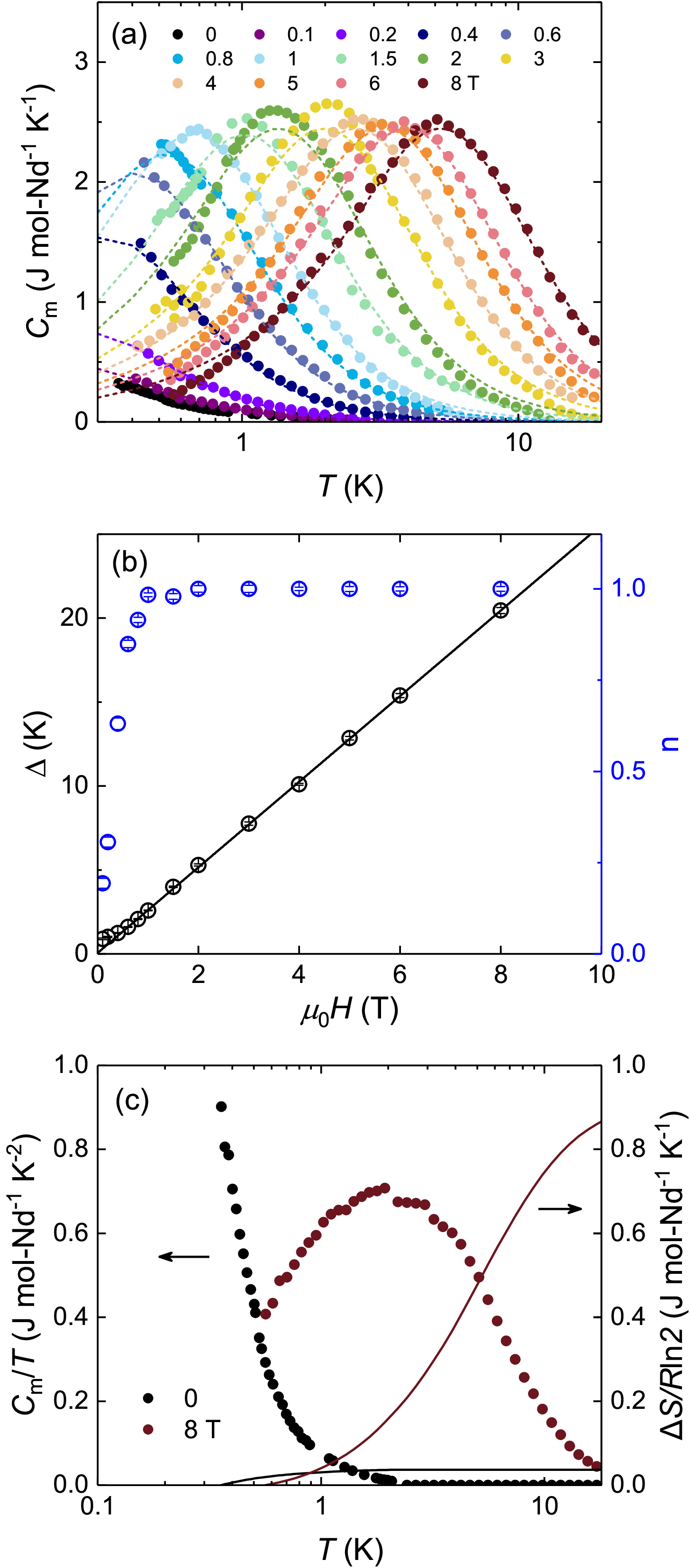}
	\caption{\label{specific2}(a) Magnetic specific heat $C_{\rm m}$ under different magnetic fields. The dashed lines are a fit to the modified two-energy Schottky model. (b) The field dependence of energy gap $\Delta$ and concentration $n$ yielded from modified Schottky fit. The black solid line is a linear fit of $\Delta$. (c) Temperature dependence of magnetic specific heat coefficient $C_{\rm m}/T$ and calculated entropy at ZF and 8~T.}
\end{figure}

 
The magnetic specific heat $C_{\rm m}$ obtained by subtracting phonon contribution from total specific heat is shown in Fig.~\ref{specific2}(a). A peak shifting to higher temperatures as magnetic field increasing is observed in $C_{\rm m}$ above 1~T, indicating a characteristic Schottky anomaly behavior, which is also reported for Ba$_6$Nd$_2$Ti$_4$O$_{17}$ single crystal previously~\cite{Song2024BTNO}. However, it is worth noting that $C_{\rm m}$ at all magnetic fields cannot be fitted well with the standard two-level Schottky function
\begin{equation}
	\label{Schottky}
	C_{\rm Sch}=nR{(\frac{\Delta}{T})}^{2}\frac{e^{-\frac{\Delta}{T}}}{{(1+e^{-\frac{\Delta}{T}})}^{2}}
\end{equation}
where $n$ is the concentration, $R$ is the molar gas constant, and $\Delta$ is the energy gap between two energy levels~\cite{tari2003specific}. We attribute that to the spatial distribution of spin directions in polycrystalline sample, Zeeman energy is different for the spins when applying a magnetic field, resulting in a distribution of energy gap $\Delta$. Thus, we assumed a uniform angular distribution of spin directions and performed numerical integration to model the two-level Schottky specific heat by the function
\begin{equation}
	\label{integ-Schottky}
	C=\int_{0}^{\pi/2} nR\frac{e^{-\Delta\cos{\theta}/T}(\Delta\cos{\theta})^2}{(1+e^{-\Delta\cos{\theta}/T})^{2}T^2}\sin{\theta} d\theta
\end{equation}
$C_{\rm m}$ is well fitted by the modified two-level Schottky model except for zero field [see dashed lines in Fig.~\ref{specific2}(a)], suggesting the two-energy-level nature at low temperatures in Ba$_6$Nd$_2$Ti$_4$O$_{17}$. The two energy levels originate form the ground-state Kramers doublet, which is dominant at low temperatures and lead to an effective spin $J_{\rm eff} = 1/2$.\par
We present the field dependence of fitting parameters $\Delta$ and $n$ in Fig.~\ref{specific2}(b). A linear fit of $\Delta$ is performed for $n \sim$1 data, the linear field dependence of energy gap $\Delta$ indicates Zeeman splitting effect caused by external applied fields. The linear fit yields $\mu$ = 2.56~$\mu_{\rm B}$ 
since the Zeeman energy is $E=\mu\mu_{\rm 0}H$, and 2.56~$\mu_{\rm B}$ nearly equals to $\mu_{\rm eff}$ = 2.54~$\mu_{\rm B}$ from magnetization measurement results. The degeneracy of ground-state doublet can be lifted by magnetic fields, but a magnetic field not large enough cannot completely lift the degeneracy and results in a concentration $n$ less than 1. The system cannot be regarded as ideal two-energy-level and single-ion at zero field and low fields. Similar behavior is reported in many Schottky anomaly results for RE-based oxide insulators~\cite{Jiang2022BYBO,Lhotel2021Yb3Ga5O12,Khatua2022Ba3RB9O18}.\par
Fig.~\ref{specific2}(c) shows the magnetic specific heat coefficient $C_{\rm m}/T$ and the corresponding entropy at ZF and 8~T. The magnetic entropy up to 20~K of 8~T nearly reach $R$ln2, which is the entropy of a spin-1/2 system. The entropy is slightly less than $R$ln2 due to the limitation of the lowest measured temperature. The ZF magnetic entropy only reach 3$\%$ of $R$ln2, suggesting considerable residual entropy below 0.4~K.
\subsection{\label{musr}$\mu$SR}
\begin{figure}[h]
	\centering
	\includegraphics[height=10cm,width=6cm]{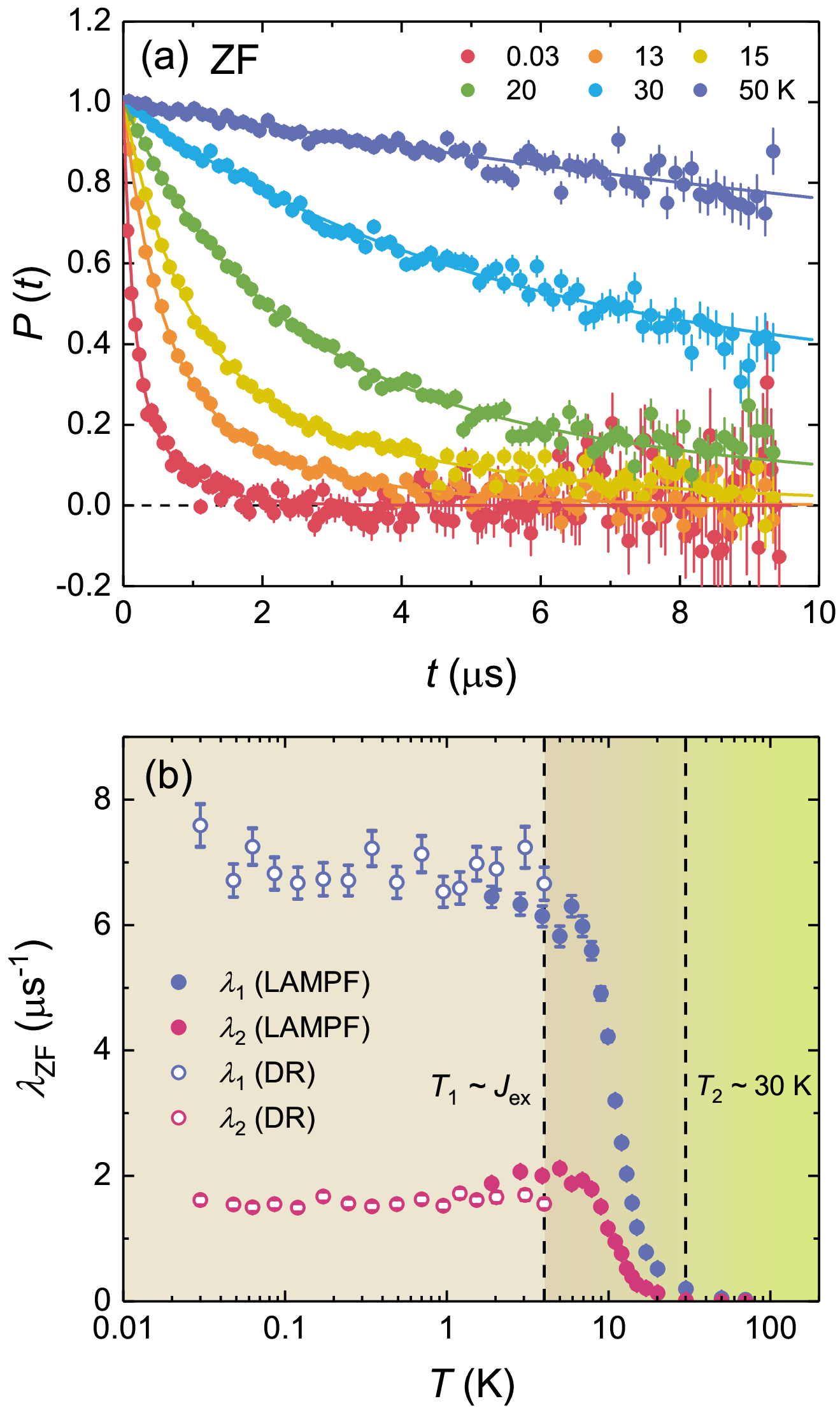}
	\caption{\label{ZF-muSR}(a) ZF-$\mu$SR time spectra of polarization at representative temperatures. The solid lines correspond to the fitting by function Eq.~\ref{ZF-ff}. (b) Temperature dependence of ZF muon spin relaxation rates $\lambda_1$ and $\lambda_2$. Open and closed circles represent data taken from the DR and the LAMPF spectrometers, respectively. $\lambda_{1,2}$ increases significantly below $T_2\sim$~30~K, and a plateau of  $\lambda_{1,2}$ is observed below $T_1\sim$~4~K.}
\end{figure}

$\mu$SR is an ideal technique to probe magnetic orderings and particularly sensitive to slow spin fluctuations~\cite{Hiller2002Muon}. We performed ZF- and LF-$\mu$SR measurements to further investigate the magnetic ground state of Ba$_6$Nd$_2$Ti$_4$O$_{17}$. ZF-$\mu$SR spectra at several representative temperatures are shown in Fig.~\ref{ZF-muSR}(a). Neither oscillations nor a drastic loss of initial asymmetry is observed down to 30~mK, ruling out any long- or short-range magnetic order~\cite{Zheng2005muSRoscillate}. The lack of polarization recovery to 1/3 suggests the absence of static random fields with Gaussian distribution~\cite{Hayano1979KT}.\par
The normalized ZF-$\mu$SR spectra with background signal subtracted are well fitted by the function
\begin{equation}
	\label{ZF-ff}
	P(t)=f_{1}e^{-{{\lambda}_{1}}t}+(1-f_{1})e^{-{{\lambda}_{2}}t}
\end{equation}
where $\lambda_1$ and $\lambda_2$ are muon spin relaxation rates, $f_1$ is the fraction of the first exponential term. $f_1$ was found to be temperature-independent and therefore fixed at the average value 0.63. The temperature dependence of the two ZF muon spin relaxation rates is shown in Fig.~\ref{ZF-muSR}(b). $\lambda_1$ and $\lambda_2$ show very similar behavior: they increase drastically below $T_{2}\sim$~30~K, and saturate at about $T_{1}\sim$~4~K, exhibiting a temperature-independent plateau as temperature further decreases down to 30~mK.

\begin{figure}[h]
	\centering
	\includegraphics[height=7.5cm,width=8.5cm]{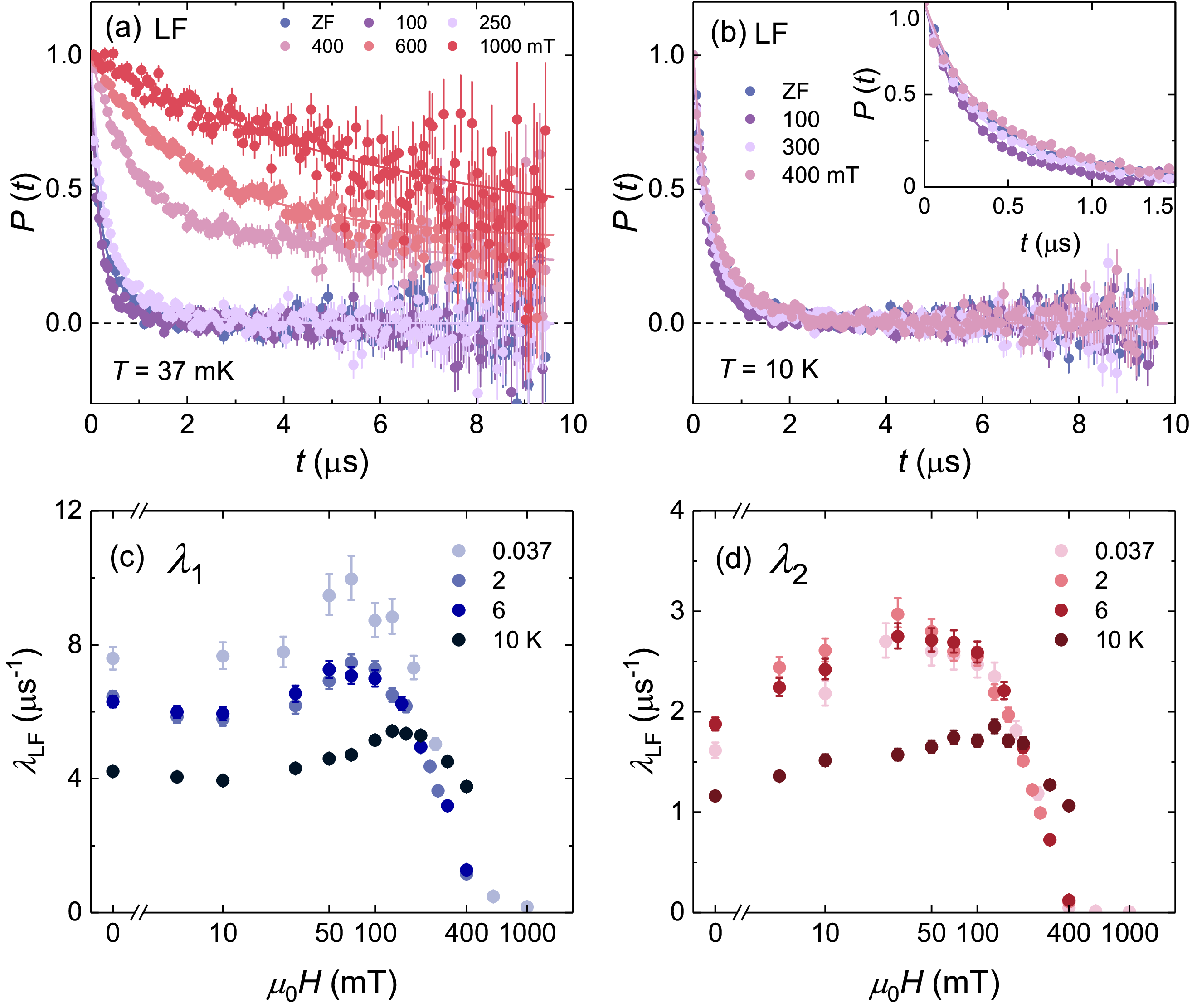}
	\caption{\label{LF-muSR} LF-$\mu$SR time spectra of polarization at representative longitudinal fields at (a) 0.037 and (b) 10~K. No decoupling phenomenon is observed under all external longitudinal fields. Inset of (b), an enlarged view of the first 1.5~$\mu$s. Field dependence of muon spin relaxation rate (c) $\lambda_1$ and (d) $\lambda_2$.}
\end{figure}

A low-temperature plateau of muon spin relaxation rate is generally regarded as the evidence of persistent dynamics and fingerprint of a correlated disordered state~\cite{Clark2019TbInO3}. To confirm the dynamics nature of spins and further investigate spin dynamics, LF-$\mu$SR measurements were carried out. Fig.~\ref{LF-muSR}(a) shows the LF-$\mu$SR spectra measured at 37~mK, the relaxation persists at various external applied fields, even longitudinal field of 1~T cannot completely decouple the muon depolarization. If the exponential depolarization function originates from a static field with Lorentzian distribution experienced by muons ~\cite{Uemura1985static}, then the distribution width $\Delta_{\rm L1,2}=\lambda_{1,2}/\gamma_{\mu}$, where $\gamma_{\mu}/2\pi$ = 135.54 MHz/T is the gyromagnetic ratio of muon. $\Delta_{\rm L1}\sim10$~mT and $\Delta_{\rm L2}\sim2$~mT are expected in this case. Generally, an external longitudinal field which is 10 times larger than the distribution width of the static internal field can fully decouple the muon depolarization~\cite{yaouanc2011muon}. 1~T is two (three) orders of magnitude larger than $\Delta_{\rm L1}$($\Delta_{\rm L2}$) and still insufficient to decouple the muon depolarization, indicating the persistent spin dynamics at 37~mK in Ba$_6$Nd$_2$Ti$_4$O$_{17}$.\par

The representative LF-$\mu$SR spectra measured at 10~K is illustrated in Fig.~\ref{LF-muSR}(b). The relaxation is hardly suppressed by an external longitudinal field of 400~mT, exhibiting a different behavior with 37~mK. The LF-$\mu$SR spectra are also well fitted by Eq.~\ref{ZF-ff}. $f_1$ is fixed to 0.63, consistent with ZF condition. We present the field dependence of muon spin relaxation rates $\lambda_1$ and $\lambda_2$ at four measured temperatures in Figs.~\ref{LF-muSR}(c) and (d). Generally, muon spin relaxation rate is suppressed monotonically with external longitudinal field increasing. However, both the field dependencies of $\lambda_1$ and $\lambda_2$ show an unusual extremum, and the extremum shifts to higher fields as temperature rises to 10~K. $\lambda_1$ and $\lambda_2$ are quenched rapidly by external longitudinal field above 100~mT below 6~K, but at 10~K, the two muon spin relaxation rates remain a sizable value comparable with ZF value even at 400~mT. The unusual field dependence of $\lambda_{1,2}$ will be discussed in detail later.\par
\section{\label{diucussions}DISCUSSIONS}
\subsection{Ising anisotropy of Nd$^{3+}$ spins}
The electron configuration of Nd$^{3+}$ is 4$f^3$. According to Hund's rule, the spin angular momentem $s=3/2$, the orbital angular momentum $L=6$, SOC leads to a total angular momentum $J=9/2$. The ground state of free Nd$^{3+}$ is ten-fold degenerate $^{4}\rm I_{9/2}$. The CEF of Ba$_6$Nd$_2$Ti$_4$O$_{17}$ splits the ten $J=9/2$ states into five Kramers doublets. The specific heat results show that only ground-state Kramers doublet exists below 200~K. Thus, the system can be treated as effective spin-1/2 at low temperatures.\par
The specific heat of single crystal Ba$_6$Nd$_2$Ti$_4$O$_{17}$ shows a perfect two-level Schottky behavior as expected for spin-1/2 system~\cite{Song2024BTNO}. While the specific heat of polycrystalline sample is described by a modified Schottky model (see Eq.~\ref{integ-Schottky}) in this work. The spin direction is assumed to follow a uniform angular distribution in the polycrystalline sample, the magnetic fields experienced by non-parallel spins are different. As a result, the energy gap of the ground-state doublet
split by the magnetic field has a distribution. However, we shall note the polycrystalline form is not the only reason for a distribution of energy gaps, another crucial reason is the Ising-like spin anisotropy in Ba$_6$Nd$_2$Ti$_4$O$_{17}$. There are many polycrystalline materials with effective 1/2 spins, the specific heat of which can be well fitted by standard two-level Schottky functions~\cite{Khatua2022Ba3RB9O18,Khatua2024BTYO,Khatua2022hyper,Ennis2024BEBO}. For a quantum spin described by the general XYZ model, the spin couples with the magnetic field in all $x$, $y$, $z$ directions. Although the spin direction has a spatial distribution in polycrystalline material, the spin coupling with magnetic field is basically the same. In contrast, Ising-like spins basically only experience the $z$ component of the magnetic field. Consequently, the Zeeman energy highly depends on the angle between the spin and the magnetic field, and results in a significant distribution of the energy gap.\par
The Ising anisotropy of Nd$^{3+}$ spins is also supported by magnetization results. The saturation value of magnetization at 2~K is $\mu_{\rm sat}$ = 1.3~$\mu_{\rm B}$, about half the value of $\mu_{\rm eff,L}$ = 2.54~$\mu_{\rm B}$, which is a sign of powder-averaged Ising spins in polycrystalline sample~\cite{Bramwell2000Isingspin}. Our results are also consistent with the statement of Ising-like spins given by electron spin resonance measurements on single crystals~\cite{Song2024BTNO}.\par
The Kramers doublet created by CEF and SOC is a dipole-octupole (DO) doublet for Nd$^{3+}$ in a triangular lattice with space group of $P6_3/mmc$ or $R\bar{3}m$~\cite{Li2016DOdoublet}. Dipolar order together with emergent octupolar order is predicted for DO doublets on triangular lattices with negligible interlayer interactions~\cite{Li2016DOdoublet}. However, no dipolar order is observed down to 30~mK in Ba$_6$Nd$_2$Ti$_4$O$_{17}$. A possible reason is that the exchange interaction between Nd$^{3+}$ spins with interlayer distance $d_1$ cannot be ignored.
\subsection{Muon stopping sites}
The fitting function for $\mu$SR spectra includes two exponential terms. We attribute that to two inequivalent muon stopping sites. The reasons are as follows: (1) The fractions of the two exponential terms are nearly temperature-independent, which is usually a sign for signals from distinct muon stopping sites in the sample~\cite{Zhu2020muSR}. (2) The fitting function for LF-$\mu$SR and ZF-$\mu$SR is the same. (3) $\lambda_1$ and $\lambda_2$ show similar temperature and field dependence. The two possible muon stopping sites are illustrated in Fig.~\ref{str}(c). 
One muon stopping site is close to the Nd$^{3+}$ triangular layers (denoted by $\mu2$), the other is located between two Nd$^{3+}$ triangular layers but close to TiO$_4$ tetrahedra or Ti$_2$O$_9$ dimers (denoted by $\mu1$). Since muons generally stop preferentially near O$^{2-}$ ions, we counted the number of O-O bonds in NdO$_6$ octahedra, and TiO$_4$ tetrahedra or Ti$_2$O$_9$ dimers respectively. If the muons stop near the first kind of O-O bonds, they can be regarded as stopping within the Nd$^{3+}$ triangular layers (at $\mu2$), whereas if the muons stop near the second kind of O-O bonds, a muon stopping site between Nd$^{3+}$ triangular layers ($\mu1$) is expected. The ratio of the two kinds of O-O bonds is 1.38, almost consistent with the ratio of two fractions of exponential terms: $0.63/0.37\approx1.70$.\par
$\lambda_1$ is several times larger than $\lambda_2$, the reason may be related with the distribution width of local magnetic field at each muon site. $\mu$2 site is located within the magnetic Nd$^{3+}$ triangular plane, whereas $\mu$1 site is located between Nd$^{3+}$ triangular planes. A relatively more uniform local magnetic field may be  experienced by muons at $\mu$2 than $\mu$1. Namely, the distribution width $\Delta_{\rm G}$ is rather small at $\mu$2 site. According to $\lambda\propto\Delta_{\rm G}^2$ for dynamic muon spin relaxation rates, a larger $\lambda_1$ (a smaller $\lambda_2$) is expected.\par
We note the above estimation is rough and can only provide some reference for the muon stopping sites. The detailed information about muon stopping sites need more density functional theory calculations.\par
\subsection{Spin dynamics at low temperatures}
Two characteristic temperatures are observed in ZF-$\mu$SR measurements. The muon spin relaxation rate $\lambda_{1,2}$ starts to rise rapidly below $T_2\sim$~30~K, and exhibit a plateau below $T_1\sim$~4~K. The energy scale of $T_1$ is of the same order of the exchange interaction $J_{\rm ex}\sim$1.2~K, suggesting the spin correlation is dominant. The low-temperature plateau of $\lambda_{1,2}$ indicates the spin fluctuation no longer slows down as temperature drops, and signals the magnetic disordered ground state with persistent dynamics, which is further confirmed by LF-$\mu$SR measurements.\par
$T_2\sim$~30~K is consistent with the Curie-Weiss temperature $\Theta_{\rm {CW},H}$ = -28.1(1)~K yielded from high-temperature fitting, suggesting $T_2$ reflects an energy scale of interaction. It is worth noting that the magnetic susceptibility deviates from high-temperature Curie-Weiss behavior at about 30~K. The deviation is generally caused by thermal depopulation from excited CEF energy levels as temperature decreases~\cite{magnetictutorial}. However, the specific heat indicates there are no excited CEF energy levels below 200~K in Ba$_6$Nd$_2$Ti$_4$O$_{17}$. The true reason for deviation is that the interaction energy of Nd$^{3+}$ becomes comparable with the thermal energy at 30~K, and the system cannot be regarded as a single-ion paramagnet any more~\cite{magnetictutorial}.\par
$T_2$ characterizes the interaction strength originated from overlap of 
Nd$^{3+}$ electron wave functions. Below 30~K, degree of freedom of the spins partly freezes, and the system can be regarded as effective spin $J_{\rm eff}$=1/2. While $T_1$ characterizes the strength of exchange interactions between the $J_{\rm eff}$=1/2 spin. Therefore, $T_1$ is one order of magnitude smaller than $T_2$.
 
\subsection{Field dependence of muon spin relaxation rates}
The internal magnetic field in Ba$_6$Nd$_2$Ti$_4$O$_{17}$ at low temperatures has been detected to be dynamical by LF-$\mu$SR measurements, but the field dependence of muon spin relaxation rates shows an unusual maximum. Level-crossing resonance (LCR) can lead to a resonance peak in the field dependence of muon spin relaxation rates at $B_{\rm res}=\omega_Q/\gamma_{\mu}$, where $\omega_Q$ is the quadrupolar frequency~\cite{kreitzman1986LCR}. LCR is established between the Zeeman interaction of the muon spin in an external magnetic field and the quadrupolar interaction of nuclei with a spin larger than 1/2~\cite{Stephen2022muon}. A resonance peak due to LCR centered at $B_{\rm {res}}\sim$~8 mT was reported for metal Cu previously~\cite{kreitzman1986LCR}. If there is a LCR peak in Ba$_6$Nd$_2$Ti$_4$O$_{17}$, $B_{\rm {res}}$ should also be of the order of several milliteslas since the quadrupole moments of Ba, Nd, Ti and Cu are of the same order (0.1~barn) and $\omega_Q$ is proportional to the quadrupole moment~\cite{yaouanc2011muon}. However, the extremum of LF muon spin relaxation rates Ba$_6$Nd$_2$Ti$_4$O$_{17}$ is observed around 100~mT. Thus, LCR can be excluded to account for the extremum.\par
For the case of random fluctuations originating from localized and uncorrelated spins, the spin dynamic autocorrelation function generally takes an exponential form, $S(t)\sim e^{-\nu t}$, where $\nu$ is the field correlation frequency. Then the muon spin relaxation rate can be expressed by the Redfield formula~\cite{Stephen2022muon}
\begin{equation}
	\label{Redfield}
	\lambda(H)=\frac{2(\gamma_{\mu}\Delta_{\rm G})^2\nu}{\nu^2+(\gamma_{\mu}\mu_{\rm 0}H)^2}
\end{equation}
in the fast fluctuation limit ($\gamma_{\mu}\Delta_{\rm G}\ll\nu$), where $\Delta_{\rm G}$ is the Gaussian distribution width of local magnetic fields, $\nu$ is the field fluctuation rate, and $\mu_{\rm 0}H$ is the external longitudinal magnetic field. $\lambda(H)$ described by the Redfield formula monotonously decreases as the external field is increased when $\Delta_{\rm G}$ and $\nu$ are nearly unchanged, which is inconsistent with the field dependence of $\lambda_{1,2}$ here.\par

We recall that the energy scale of exchange interactions is about 1~K for Ba$_6$Nd$_2$Ti$_4$O$_{17}$. At $T$ = 10~K, thermal fluctuations are relatively strong, and the exchange interactions are negligible. The extremum of $\lambda_{1,2}$ persists at temperature up to 10~K, suggesting it cannot result from exchange interactions or spin correlation. The extremum of $\lambda_{1,2}$ may be due to the change in the distribution width of local fields $\Delta_{\rm G}$. Since the field coupling of Ising spins are highly anisotropic and the sample is polycrystalline, the longitudinal field dependence of  $\Delta_{\rm G}$ may be significant to drive $\lambda(H)$ no longer monotonous.\par

Both $\lambda_1$ and $\lambda_2$ remain a sizable value at 10~K under 400~mT. The behavior of $\lambda_{1,2}$ hardly suppressed by the external longitudinal field is reminiscent of the field-independent term observed in LF muon spin relaxation rate for YbZnGaO$_4$, which is due to rapidly fluctuating localized spins~\cite{Pratt2022YZGO}. Since thermal fluctuations are dominant at 10~K in Ba$_6$Nd$_2$Ti$_4$O$_{17}$, the field-robust LF muon spin relaxation rates are highly likely a result of spin fluctuation due to thermal effect. We can estimate the spin fluctuation rate $\nu=\sqrt{z}J_0s/h\sim$~3$\times$10$^4$~MHz in the high-temperature limit, where $z$ = 6 is the coordination number of spins for a triangular lattice~\cite{Li2016musrYMGO,Uemura1994nu}. Then using Eq.~\ref{Redfield} with $H$ = 0 and $\lambda\sim$~1~MHz (of the order of  ZF values of $\lambda_1$ and $\lambda_2$), the distribution width of local field $\Delta_{\rm G}$ is estimated to be of the order of 10$^2$~mT. The fast fluctuation limit is satisfied 
($\gamma_{\mu}\Delta_{\rm G}\sim$~10$^2$~MHz$\ll\nu\sim$~10$^4$~MHz), so the estimation above is valid. The estimated $\nu$ and $\Delta_{\rm G}$ are comparable to the values reported for YbMgGaO$_4$ before ($\nu\sim$~4$\times$10$^4$~MHz, $\Delta_{\rm G}\sim$~70~mT)~\cite{Li2016musrYMGO}.\par

\section{\label{Conc}CONCLUSIONS}
In conclusion, the antiferromagnet polycrystalline Ba$_6$Nd$_2$Ti$_4$O$_{17}$ with a disorder-free triangular lattice has been synthesized and studied. The absence of long-range magnetic order and spin freezing is confirmed down to 30~mK by ZF-$\mu$SR measurements. The negative Curie-Weiss temperature $\Theta_{\rm {CW}}$ = -1.8~K yielded form magnetic susceptibility measurements indicates the antiferromagnetic exchange interaction of Nd$^{3+}$ spins. The low-temperature magnetic specific heat is well fitted by the modified two-level Schottky model, suggesting the system has an effective Ising spin $J_{\rm eff} = 1/2$ at low temperatures due to ground-state Kramers doublet. The Ising spin anisotropy is also supported by the effective magnetic moment yielded from magnetization measurements. The low-temperature plateau of muon spin relaxation rates in ZF-$\mu$SR indicates a quantum disordered ground state. LF-$\mu$SR measurements further reveal the persistent spin dynamics at low temperature down to 37 mK.\par
Our work shows the intriguing behavior of a triangular-lattice antiferromagnet with effective-1/2 Ising spins. We provide an excellent example which remains magnetically disordered down to $\left| \Theta_{\rm {CW}} \right|/10$ with persistent spin dynamics. However, more experiments are required to investigate the spin excitations at low temperatures and check the ground state, especially the experiments on single crystals.

\begin{acknowledgments}
We are grateful to G. D. Morris, B. Hitti, and D. Arsenau of the TRIUMF CMMS for assistance during the experiments. We thank G. Chen, Y. Wan and Y. Wang for discussions. This research was funded by the National Key Research and Development Program of China, No. 2022YFA1402203, the National Natural Science Foundations of China, No. 12174065,  and the Shanghai Municipal Science and Technology Major Project (Grant No. 2019SHZDZX01). Research at CSU-Fresno was supported by NSF DMR-1905636.
\end{acknowledgments}

\begin{thebibliography}{53}%
\makeatletter
\providecommand \@ifxundefined [1]{%
 \@ifx{#1\undefined}
}%
\providecommand \@ifnum [1]{%
 \ifnum #1\expandafter \@firstoftwo
 \else \expandafter \@secondoftwo
 \fi
}%
\providecommand \@ifx [1]{%
 \ifx #1\expandafter \@firstoftwo
 \else \expandafter \@secondoftwo
 \fi
}%
\providecommand \natexlab [1]{#1}%
\providecommand \enquote  [1]{``#1''}%
\providecommand \bibnamefont  [1]{#1}%
\providecommand \bibfnamefont [1]{#1}%
\providecommand \citenamefont [1]{#1}%
\providecommand \href@noop [0]{\@secondoftwo}%
\providecommand \href [0]{\begingroup \@sanitize@url \@href}%
\providecommand \@href[1]{\@@startlink{#1}\@@href}%
\providecommand \@@href[1]{\endgroup#1\@@endlink}%
\providecommand \@sanitize@url [0]{\catcode `\\12\catcode `\$12\catcode
  `\&12\catcode `\#12\catcode `\^12\catcode `\_12\catcode `\%12\relax}%
\providecommand \@@startlink[1]{}%
\providecommand \@@endlink[0]{}%
\providecommand \url  [0]{\begingroup\@sanitize@url \@url }%
\providecommand \@url [1]{\endgroup\@href {#1}{\urlprefix }}%
\providecommand \urlprefix  [0]{URL }%
\providecommand \Eprint [0]{\href }%
\providecommand \doibase [0]{https://doi.org/}%
\providecommand \selectlanguage [0]{\@gobble}%
\providecommand \bibinfo  [0]{\@secondoftwo}%
\providecommand \bibfield  [0]{\@secondoftwo}%
\providecommand \translation [1]{[#1]}%
\providecommand \BibitemOpen [0]{}%
\providecommand \bibitemStop [0]{}%
\providecommand \bibitemNoStop [0]{.\EOS\space}%
\providecommand \EOS [0]{\spacefactor3000\relax}%
\providecommand \BibitemShut  [1]{\csname bibitem#1\endcsname}%
\let\auto@bib@innerbib\@empty
\bibitem [{\citenamefont {Balents}(2010)}]{balents2010QSL}%
  \BibitemOpen
  \bibfield  {author} {\bibinfo {author} {\bibfnamefont {L.}~\bibnamefont
  {Balents}},\ }\href {https://doi.org/10.1038/nature08917} {\bibfield
  {journal} {\bibinfo  {journal} {Nature}\ }\textbf {\bibinfo {volume} {464}},\
  \bibinfo {pages} {199} (\bibinfo {year} {2010})}\BibitemShut {NoStop}%
\bibitem [{\citenamefont {Anderson}(1987)}]{anderson1987RVB}%
  \BibitemOpen
  \bibfield  {author} {\bibinfo {author} {\bibfnamefont {P.~W.}\ \bibnamefont
  {Anderson}},\ }\href {https://doi.org/10.1126/science.235.4793.1196}
  {\bibfield  {journal} {\bibinfo  {journal} {Science}\ }\textbf {\bibinfo
  {volume} {235}},\ \bibinfo {pages} {1196} (\bibinfo {year}
  {1987})}\BibitemShut {NoStop}%
\bibitem [{\citenamefont {Wen}(2002)}]{wen2002QSL}%
  \BibitemOpen
  \bibfield  {author} {\bibinfo {author} {\bibfnamefont {X.-G.}\ \bibnamefont
  {Wen}},\ }\href {https://doi.org/10.1103/PhysRevB.65.165113} {\bibfield
  {journal} {\bibinfo  {journal} {Phys. Rev. B}\ }\textbf {\bibinfo {volume}
  {65}},\ \bibinfo {pages} {165113} (\bibinfo {year} {2002})}\BibitemShut
  {NoStop}%
\bibitem [{\citenamefont {Takagi}\ \emph {et~al.}(2019)\citenamefont {Takagi},
  \citenamefont {Takayama}, \citenamefont {Jackeli}, \citenamefont
  {Khaliullin},\ and\ \citenamefont {Nagler}}]{Takagi2019KQSL}%
  \BibitemOpen
  \bibfield  {author} {\bibinfo {author} {\bibfnamefont {H.}~\bibnamefont
  {Takagi}}, \bibinfo {author} {\bibfnamefont {T.}~\bibnamefont {Takayama}},
  \bibinfo {author} {\bibfnamefont {G.}~\bibnamefont {Jackeli}}, \bibinfo
  {author} {\bibfnamefont {G.}~\bibnamefont {Khaliullin}},\ and\ \bibinfo
  {author} {\bibfnamefont {S.~E.}\ \bibnamefont {Nagler}},\ }\href
  {https://doi.org/10.1038/s42254-019-0038-2} {\bibfield  {journal} {\bibinfo
  {journal} {Nat. Rev. Phys.}\ }\textbf {\bibinfo {volume} {1}},\ \bibinfo
  {pages} {264} (\bibinfo {year} {2019})}\BibitemShut {NoStop}%
\bibitem [{\citenamefont {Wen}(1989)}]{Wen1989frac}%
  \BibitemOpen
  \bibfield  {author} {\bibinfo {author} {\bibfnamefont {X.~G.}\ \bibnamefont
  {Wen}},\ }\href {https://doi.org/10.1103/PhysRevB.39.7223} {\bibfield
  {journal} {\bibinfo  {journal} {Phys. Rev. B}\ }\textbf {\bibinfo {volume}
  {39}},\ \bibinfo {pages} {7223} (\bibinfo {year} {1989})}\BibitemShut
  {NoStop}%
\bibitem [{\citenamefont {Wen}(1991)}]{Wen1991frac}%
  \BibitemOpen
  \bibfield  {author} {\bibinfo {author} {\bibfnamefont {X.~G.}\ \bibnamefont
  {Wen}},\ }\href {https://doi.org/10.1103/PhysRevB.44.2664} {\bibfield
  {journal} {\bibinfo  {journal} {Phys. Rev. B}\ }\textbf {\bibinfo {volume}
  {44}},\ \bibinfo {pages} {2664} (\bibinfo {year} {1991})}\BibitemShut
  {NoStop}%
\bibitem [{\citenamefont {Anderson}(1973)}]{Anderson1973QSL}%
  \BibitemOpen
  \bibfield  {author} {\bibinfo {author} {\bibfnamefont {P.}~\bibnamefont
  {Anderson}},\ }\href
  {https://doi.org/https://doi.org/10.1016/0025-5408(73)90167-0} {\bibfield
  {journal} {\bibinfo  {journal} {Mater. Res. Bull.}\ }\textbf {\bibinfo
  {volume} {8}},\ \bibinfo {pages} {153} (\bibinfo {year} {1973})}\BibitemShut
  {NoStop}%
\bibitem [{\citenamefont {Huse}\ and\ \citenamefont
  {Elser}(1988)}]{Huse1988Hesienberg}%
  \BibitemOpen
  \bibfield  {author} {\bibinfo {author} {\bibfnamefont {D.~A.}\ \bibnamefont
  {Huse}}\ and\ \bibinfo {author} {\bibfnamefont {V.}~\bibnamefont {Elser}},\
  }\href {https://doi.org/10.1103/PhysRevLett.60.2531} {\bibfield  {journal}
  {\bibinfo  {journal} {Phys. Rev. Lett.}\ }\textbf {\bibinfo {volume} {60}},\
  \bibinfo {pages} {2531} (\bibinfo {year} {1988})}\BibitemShut {NoStop}%
\bibitem [{\citenamefont {Iqbal}\ \emph {et~al.}(2016)\citenamefont {Iqbal},
  \citenamefont {Hu}, \citenamefont {Thomale}, \citenamefont {Poilblanc},\ and\
  \citenamefont {Becca}}]{Iqbal2016J1J2}%
  \BibitemOpen
  \bibfield  {author} {\bibinfo {author} {\bibfnamefont {Y.}~\bibnamefont
  {Iqbal}}, \bibinfo {author} {\bibfnamefont {W.-J.}\ \bibnamefont {Hu}},
  \bibinfo {author} {\bibfnamefont {R.}~\bibnamefont {Thomale}}, \bibinfo
  {author} {\bibfnamefont {D.}~\bibnamefont {Poilblanc}},\ and\ \bibinfo
  {author} {\bibfnamefont {F.}~\bibnamefont {Becca}},\ }\href
  {https://doi.org/10.1103/PhysRevB.93.144411} {\bibfield  {journal} {\bibinfo
  {journal} {Phys. Rev. B}\ }\textbf {\bibinfo {volume} {93}},\ \bibinfo
  {pages} {144411} (\bibinfo {year} {2016})}\BibitemShut {NoStop}%
\bibitem [{\citenamefont {Zhu}\ and\ \citenamefont
  {White}(2015)}]{Zhu2015J1J2}%
  \BibitemOpen
  \bibfield  {author} {\bibinfo {author} {\bibfnamefont {Z.}~\bibnamefont
  {Zhu}}\ and\ \bibinfo {author} {\bibfnamefont {S.~R.}\ \bibnamefont
  {White}},\ }\href {https://doi.org/10.1103/PhysRevB.92.041105} {\bibfield
  {journal} {\bibinfo  {journal} {Phys. Rev. B}\ }\textbf {\bibinfo {volume}
  {92}},\ \bibinfo {pages} {041105} (\bibinfo {year} {2015})}\BibitemShut
  {NoStop}%
\bibitem [{\citenamefont {Luo}\ \emph {et~al.}(2017)\citenamefont {Luo},
  \citenamefont {Hu}, \citenamefont {Xi}, \citenamefont {Zhao},\ and\
  \citenamefont {Wang}}]{Luo2017anisotropy}%
  \BibitemOpen
  \bibfield  {author} {\bibinfo {author} {\bibfnamefont {Q.}~\bibnamefont
  {Luo}}, \bibinfo {author} {\bibfnamefont {S.}~\bibnamefont {Hu}}, \bibinfo
  {author} {\bibfnamefont {B.}~\bibnamefont {Xi}}, \bibinfo {author}
  {\bibfnamefont {J.}~\bibnamefont {Zhao}},\ and\ \bibinfo {author}
  {\bibfnamefont {X.}~\bibnamefont {Wang}},\ }\href
  {https://doi.org/10.1103/PhysRevB.95.165110} {\bibfield  {journal} {\bibinfo
  {journal} {Phys. Rev. B}\ }\textbf {\bibinfo {volume} {95}},\ \bibinfo
  {pages} {165110} (\bibinfo {year} {2017})}\BibitemShut {NoStop}%
\bibitem [{\citenamefont {Motrunich}(2005)}]{Motrunich2005anisotropy}%
  \BibitemOpen
  \bibfield  {author} {\bibinfo {author} {\bibfnamefont {O.~I.}\ \bibnamefont
  {Motrunich}},\ }\href {https://doi.org/10.1103/PhysRevB.72.045105} {\bibfield
   {journal} {\bibinfo  {journal} {Phys. Rev. B}\ }\textbf {\bibinfo {volume}
  {72}},\ \bibinfo {pages} {045105} (\bibinfo {year} {2005})}\BibitemShut
  {NoStop}%
\bibitem [{\citenamefont {Yamamoto}\ \emph {et~al.}(2014)\citenamefont
  {Yamamoto}, \citenamefont {Marmorini},\ and\ \citenamefont
  {Danshita}}]{Yamamoto2014anisotropy}%
  \BibitemOpen
  \bibfield  {author} {\bibinfo {author} {\bibfnamefont {D.}~\bibnamefont
  {Yamamoto}}, \bibinfo {author} {\bibfnamefont {G.}~\bibnamefont
  {Marmorini}},\ and\ \bibinfo {author} {\bibfnamefont {I.}~\bibnamefont
  {Danshita}},\ }\href {https://doi.org/10.1103/PhysRevLett.112.127203}
  {\bibfield  {journal} {\bibinfo  {journal} {Phys. Rev. Lett.}\ }\textbf
  {\bibinfo {volume} {112}},\ \bibinfo {pages} {127203} (\bibinfo {year}
  {2014})}\BibitemShut {NoStop}%
\bibitem [{\citenamefont {Li}\ \emph {et~al.}(2016{\natexlab{a}})\citenamefont
  {Li}, \citenamefont {Wang},\ and\ \citenamefont
  {Chen}}]{Li2016Kramerstheory}%
  \BibitemOpen
  \bibfield  {author} {\bibinfo {author} {\bibfnamefont {Y.-D.}\ \bibnamefont
  {Li}}, \bibinfo {author} {\bibfnamefont {X.}~\bibnamefont {Wang}},\ and\
  \bibinfo {author} {\bibfnamefont {G.}~\bibnamefont {Chen}},\ }\href
  {https://doi.org/10.1103/PhysRevB.94.035107} {\bibfield  {journal} {\bibinfo
  {journal} {Phys. Rev. B}\ }\textbf {\bibinfo {volume} {94}},\ \bibinfo
  {pages} {035107} (\bibinfo {year} {2016}{\natexlab{a}})}\BibitemShut
  {NoStop}%
\bibitem [{\citenamefont {Witczak-Krempa}\ \emph {et~al.}(2014)\citenamefont
  {Witczak-Krempa}, \citenamefont {Chen}, \citenamefont {Kim},\ and\
  \citenamefont {Balents}}]{William2013SOC}%
  \BibitemOpen
  \bibfield  {author} {\bibinfo {author} {\bibfnamefont {W.}~\bibnamefont
  {Witczak-Krempa}}, \bibinfo {author} {\bibfnamefont {G.}~\bibnamefont
  {Chen}}, \bibinfo {author} {\bibfnamefont {Y.~B.}\ \bibnamefont {Kim}},\ and\
  \bibinfo {author} {\bibfnamefont {L.}~\bibnamefont {Balents}},\ }\href
  {https://doi.org/https://doi.org/10.1146/annurev-conmatphys-020911-125138}
  {\bibfield  {journal} {\bibinfo  {journal} {Annu. Rev. Condens. Matter
  Phys.}\ }\textbf {\bibinfo {volume} {5}},\ \bibinfo {pages} {57} (\bibinfo
  {year} {2014})}\BibitemShut {NoStop}%
\bibitem [{\citenamefont {Li}\ \emph {et~al.}(2015)\citenamefont {Li},
  \citenamefont {Liao}, \citenamefont {Zhang}, \citenamefont {Li},
  \citenamefont {Jin}, \citenamefont {Ling}, \citenamefont {Zhang},
  \citenamefont {Zou}, \citenamefont {Pi}, \citenamefont {Yang}, \citenamefont
  {Wang}, \citenamefont {Wu},\ and\ \citenamefont {Zhang}}]{li2015YMGO}%
  \BibitemOpen
  \bibfield  {author} {\bibinfo {author} {\bibfnamefont {Y.}~\bibnamefont
  {Li}}, \bibinfo {author} {\bibfnamefont {H.}~\bibnamefont {Liao}}, \bibinfo
  {author} {\bibfnamefont {Z.}~\bibnamefont {Zhang}}, \bibinfo {author}
  {\bibfnamefont {S.}~\bibnamefont {Li}}, \bibinfo {author} {\bibfnamefont
  {F.}~\bibnamefont {Jin}}, \bibinfo {author} {\bibfnamefont {L.}~\bibnamefont
  {Ling}}, \bibinfo {author} {\bibfnamefont {L.}~\bibnamefont {Zhang}},
  \bibinfo {author} {\bibfnamefont {Y.}~\bibnamefont {Zou}}, \bibinfo {author}
  {\bibfnamefont {L.}~\bibnamefont {Pi}}, \bibinfo {author} {\bibfnamefont
  {Z.}~\bibnamefont {Yang}}, \bibinfo {author} {\bibfnamefont {J.}~\bibnamefont
  {Wang}}, \bibinfo {author} {\bibfnamefont {Z.}~\bibnamefont {Wu}},\ and\
  \bibinfo {author} {\bibfnamefont {Q.}~\bibnamefont {Zhang}},\ }\href
  {https://doi.org/10.1038/srep16419} {\bibfield  {journal} {\bibinfo
  {journal} {Sci. Rep.}\ }\textbf {\bibinfo {volume} {5}},\ \bibinfo {pages}
  {16419} (\bibinfo {year} {2015})}\BibitemShut {NoStop}%
\bibitem [{\citenamefont {Shen}\ \emph {et~al.}(2016)\citenamefont {Shen},
  \citenamefont {Li}, \citenamefont {Wo}, \citenamefont {Li}, \citenamefont
  {Shen}, \citenamefont {Pan}, \citenamefont {Wang}, \citenamefont {Walker},
  \citenamefont {Steffens}, \citenamefont {Boehm}, \citenamefont {Hao},
  \citenamefont {Quintero-Castro}, \citenamefont {Harriger}, \citenamefont
  {Frontzek}, \citenamefont {Hao}, \citenamefont {Meng}, \citenamefont {Zhang},
  \citenamefont {Chen},\ and\ \citenamefont {Zhao}}]{shen2016YMGO}%
  \BibitemOpen
  \bibfield  {author} {\bibinfo {author} {\bibfnamefont {Y.}~\bibnamefont
  {Shen}}, \bibinfo {author} {\bibfnamefont {Y.-D.}\ \bibnamefont {Li}},
  \bibinfo {author} {\bibfnamefont {H.}~\bibnamefont {Wo}}, \bibinfo {author}
  {\bibfnamefont {Y.}~\bibnamefont {Li}}, \bibinfo {author} {\bibfnamefont
  {S.}~\bibnamefont {Shen}}, \bibinfo {author} {\bibfnamefont {B.}~\bibnamefont
  {Pan}}, \bibinfo {author} {\bibfnamefont {Q.}~\bibnamefont {Wang}}, \bibinfo
  {author} {\bibfnamefont {H.~C.}\ \bibnamefont {Walker}}, \bibinfo {author}
  {\bibfnamefont {P.}~\bibnamefont {Steffens}}, \bibinfo {author}
  {\bibfnamefont {M.}~\bibnamefont {Boehm}}, \bibinfo {author} {\bibfnamefont
  {Y.}~\bibnamefont {Hao}}, \bibinfo {author} {\bibfnamefont {D.~L.}\
  \bibnamefont {Quintero-Castro}}, \bibinfo {author} {\bibfnamefont {L.~W.}\
  \bibnamefont {Harriger}}, \bibinfo {author} {\bibfnamefont {M.~D.}\
  \bibnamefont {Frontzek}}, \bibinfo {author} {\bibfnamefont {L.}~\bibnamefont
  {Hao}}, \bibinfo {author} {\bibfnamefont {S.}~\bibnamefont {Meng}}, \bibinfo
  {author} {\bibfnamefont {Q.}~\bibnamefont {Zhang}}, \bibinfo {author}
  {\bibfnamefont {G.}~\bibnamefont {Chen}},\ and\ \bibinfo {author}
  {\bibfnamefont {J.}~\bibnamefont {Zhao}},\ }\href
  {https://doi.org/10.1038/nature20614} {\bibfield  {journal} {\bibinfo
  {journal} {Nature}\ }\textbf {\bibinfo {volume} {540}},\ \bibinfo {pages}
  {559} (\bibinfo {year} {2016})}\BibitemShut {NoStop}%
\bibitem [{\citenamefont {Li}\ \emph {et~al.}(2017)\citenamefont {Li},
  \citenamefont {Lu},\ and\ \citenamefont {Chen}}]{li2017YMGO}%
  \BibitemOpen
  \bibfield  {author} {\bibinfo {author} {\bibfnamefont {Y.-D.}\ \bibnamefont
  {Li}}, \bibinfo {author} {\bibfnamefont {Y.-M.}\ \bibnamefont {Lu}},\ and\
  \bibinfo {author} {\bibfnamefont {G.}~\bibnamefont {Chen}},\ }\href
  {https://doi.org/10.1103/PhysRevB.96.054445} {\bibfield  {journal} {\bibinfo
  {journal} {Phys. Rev. B}\ }\textbf {\bibinfo {volume} {96}},\ \bibinfo
  {pages} {054445} (\bibinfo {year} {2017})}\BibitemShut {NoStop}%
\bibitem [{\citenamefont {Ding}\ \emph {et~al.}(2019)\citenamefont {Ding},
  \citenamefont {Manuel}, \citenamefont {Bachus}, \citenamefont {Gru\ss{}ler},
  \citenamefont {Gegenwart}, \citenamefont {Singleton}, \citenamefont
  {Johnson}, \citenamefont {Walker}, \citenamefont {Adroja}, \citenamefont
  {Hillier},\ and\ \citenamefont {Tsirlin}}]{Ding2019NaYbO2}%
  \BibitemOpen
  \bibfield  {author} {\bibinfo {author} {\bibfnamefont {L.}~\bibnamefont
  {Ding}}, \bibinfo {author} {\bibfnamefont {P.}~\bibnamefont {Manuel}},
  \bibinfo {author} {\bibfnamefont {S.}~\bibnamefont {Bachus}}, \bibinfo
  {author} {\bibfnamefont {F.}~\bibnamefont {Gru\ss{}ler}}, \bibinfo {author}
  {\bibfnamefont {P.}~\bibnamefont {Gegenwart}}, \bibinfo {author}
  {\bibfnamefont {J.}~\bibnamefont {Singleton}}, \bibinfo {author}
  {\bibfnamefont {R.~D.}\ \bibnamefont {Johnson}}, \bibinfo {author}
  {\bibfnamefont {H.~C.}\ \bibnamefont {Walker}}, \bibinfo {author}
  {\bibfnamefont {D.~T.}\ \bibnamefont {Adroja}}, \bibinfo {author}
  {\bibfnamefont {A.~D.}\ \bibnamefont {Hillier}},\ and\ \bibinfo {author}
  {\bibfnamefont {A.~A.}\ \bibnamefont {Tsirlin}},\ }\href
  {https://doi.org/10.1103/PhysRevB.100.144432} {\bibfield  {journal} {\bibinfo
   {journal} {Phys. Rev. B}\ }\textbf {\bibinfo {volume} {100}},\ \bibinfo
  {pages} {144432} (\bibinfo {year} {2019})}\BibitemShut {NoStop}%
\bibitem [{\citenamefont {Baenitz}\ \emph {et~al.}(2018)\citenamefont
  {Baenitz}, \citenamefont {Schlender}, \citenamefont {Sichelschmidt},
  \citenamefont {Onykiienko}, \citenamefont {Zangeneh}, \citenamefont
  {Ranjith}, \citenamefont {Sarkar}, \citenamefont {Hozoi}, \citenamefont
  {Walker}, \citenamefont {Orain}, \citenamefont {Yasuoka}, \citenamefont
  {van~den Brink}, \citenamefont {Klauss}, \citenamefont {Inosov},\ and\
  \citenamefont {Doert}}]{Baenitz2018NaYbS2}%
  \BibitemOpen
  \bibfield  {author} {\bibinfo {author} {\bibfnamefont {M.}~\bibnamefont
  {Baenitz}}, \bibinfo {author} {\bibfnamefont {P.}~\bibnamefont {Schlender}},
  \bibinfo {author} {\bibfnamefont {J.}~\bibnamefont {Sichelschmidt}}, \bibinfo
  {author} {\bibfnamefont {Y.~A.}\ \bibnamefont {Onykiienko}}, \bibinfo
  {author} {\bibfnamefont {Z.}~\bibnamefont {Zangeneh}}, \bibinfo {author}
  {\bibfnamefont {K.~M.}\ \bibnamefont {Ranjith}}, \bibinfo {author}
  {\bibfnamefont {R.}~\bibnamefont {Sarkar}}, \bibinfo {author} {\bibfnamefont
  {L.}~\bibnamefont {Hozoi}}, \bibinfo {author} {\bibfnamefont {H.~C.}\
  \bibnamefont {Walker}}, \bibinfo {author} {\bibfnamefont {J.-C.}\
  \bibnamefont {Orain}}, \bibinfo {author} {\bibfnamefont {H.}~\bibnamefont
  {Yasuoka}}, \bibinfo {author} {\bibfnamefont {J.}~\bibnamefont {van~den
  Brink}}, \bibinfo {author} {\bibfnamefont {H.~H.}\ \bibnamefont {Klauss}},
  \bibinfo {author} {\bibfnamefont {D.~S.}\ \bibnamefont {Inosov}},\ and\
  \bibinfo {author} {\bibfnamefont {T.}~\bibnamefont {Doert}},\ }\href
  {https://doi.org/10.1103/PhysRevB.98.220409} {\bibfield  {journal} {\bibinfo
  {journal} {Phys. Rev. B}\ }\textbf {\bibinfo {volume} {98}},\ \bibinfo
  {pages} {220409} (\bibinfo {year} {2018})}\BibitemShut {NoStop}%
\bibitem [{\citenamefont {Zhu}\ \emph {et~al.}(2023)\citenamefont {Zhu},
  \citenamefont {Pan}, \citenamefont {Nie}, \citenamefont {Ni}, \citenamefont
  {Yang}, \citenamefont {Chen}, \citenamefont {Jiang}, \citenamefont {Huang},
  \citenamefont {Cheng}, \citenamefont {Yu}, \citenamefont {Miao},
  \citenamefont {Hillier}, \citenamefont {Chen}, \citenamefont {Wu},
  \citenamefont {Zhou}, \citenamefont {Li},\ and\ \citenamefont
  {Shu}}]{Zhu2023NaYbSe2}%
  \BibitemOpen
  \bibfield  {author} {\bibinfo {author} {\bibfnamefont {Z.}~\bibnamefont
  {Zhu}}, \bibinfo {author} {\bibfnamefont {B.}~\bibnamefont {Pan}}, \bibinfo
  {author} {\bibfnamefont {L.}~\bibnamefont {Nie}}, \bibinfo {author}
  {\bibfnamefont {J.}~\bibnamefont {Ni}}, \bibinfo {author} {\bibfnamefont
  {Y.}~\bibnamefont {Yang}}, \bibinfo {author} {\bibfnamefont {C.}~\bibnamefont
  {Chen}}, \bibinfo {author} {\bibfnamefont {C.}~\bibnamefont {Jiang}},
  \bibinfo {author} {\bibfnamefont {Y.}~\bibnamefont {Huang}}, \bibinfo
  {author} {\bibfnamefont {E.}~\bibnamefont {Cheng}}, \bibinfo {author}
  {\bibfnamefont {Y.}~\bibnamefont {Yu}}, \bibinfo {author} {\bibfnamefont
  {J.}~\bibnamefont {Miao}}, \bibinfo {author} {\bibfnamefont {A.~D.}\
  \bibnamefont {Hillier}}, \bibinfo {author} {\bibfnamefont {X.}~\bibnamefont
  {Chen}}, \bibinfo {author} {\bibfnamefont {T.}~\bibnamefont {Wu}}, \bibinfo
  {author} {\bibfnamefont {Y.}~\bibnamefont {Zhou}}, \bibinfo {author}
  {\bibfnamefont {S.}~\bibnamefont {Li}},\ and\ \bibinfo {author}
  {\bibfnamefont {L.}~\bibnamefont {Shu}},\ }\href
  {https://doi.org/https://doi.org/10.1016/j.xinn.2023.100459} {\bibfield
  {journal} {\bibinfo  {journal} {Innovation}\ }\textbf {\bibinfo {volume}
  {4}},\ \bibinfo {pages} {100459} (\bibinfo {year} {2023})}\BibitemShut
  {NoStop}%
\bibitem [{\citenamefont {Arh}\ \emph {et~al.}(2022)\citenamefont {Arh},
  \citenamefont {Sana}, \citenamefont {Pregelj}, \citenamefont {Khuntia},
  \citenamefont {Jagličić}, \citenamefont {Le}, \citenamefont {Biswas},
  \citenamefont {Manuel}, \citenamefont {Mangin-Thro}, \citenamefont
  {Ozarowski},\ and\ \citenamefont {Zorko}}]{Arh2022NdTa7O19}%
  \BibitemOpen
  \bibfield  {author} {\bibinfo {author} {\bibfnamefont {T.}~\bibnamefont
  {Arh}}, \bibinfo {author} {\bibfnamefont {B.}~\bibnamefont {Sana}}, \bibinfo
  {author} {\bibfnamefont {M.}~\bibnamefont {Pregelj}}, \bibinfo {author}
  {\bibfnamefont {P.}~\bibnamefont {Khuntia}}, \bibinfo {author} {\bibfnamefont
  {Z.}~\bibnamefont {Jagličić}}, \bibinfo {author} {\bibfnamefont {M.~D.}\
  \bibnamefont {Le}}, \bibinfo {author} {\bibfnamefont {P.~K.}\ \bibnamefont
  {Biswas}}, \bibinfo {author} {\bibfnamefont {P.}~\bibnamefont {Manuel}},
  \bibinfo {author} {\bibfnamefont {L.}~\bibnamefont {Mangin-Thro}}, \bibinfo
  {author} {\bibfnamefont {A.}~\bibnamefont {Ozarowski}},\ and\ \bibinfo
  {author} {\bibfnamefont {A.}~\bibnamefont {Zorko}},\ }\href
  {https://doi.org/10.1038/s41563-021-01169-y} {\bibfield  {journal} {\bibinfo
  {journal} {Nat. Mater.}\ }\textbf {\bibinfo {volume} {21}},\ \bibinfo {pages}
  {416} (\bibinfo {year} {2022})}\BibitemShut {NoStop}%
\bibitem [{\citenamefont {Khatua}\ \emph {et~al.}(2023)\citenamefont {Khatua},
  \citenamefont {Sana}, \citenamefont {Zorko}, \citenamefont {Gomilšek},
  \citenamefont {Sethupathi}, \citenamefont {Rao}, \citenamefont {Baenitz},
  \citenamefont {Schmidt},\ and\ \citenamefont {Khuntia}}]{Khatua2023review}%
  \BibitemOpen
  \bibfield  {author} {\bibinfo {author} {\bibfnamefont {J.}~\bibnamefont
  {Khatua}}, \bibinfo {author} {\bibfnamefont {B.}~\bibnamefont {Sana}},
  \bibinfo {author} {\bibfnamefont {A.}~\bibnamefont {Zorko}}, \bibinfo
  {author} {\bibfnamefont {M.}~\bibnamefont {Gomilšek}}, \bibinfo {author}
  {\bibfnamefont {K.}~\bibnamefont {Sethupathi}}, \bibinfo {author}
  {\bibfnamefont {M.~R.}\ \bibnamefont {Rao}}, \bibinfo {author} {\bibfnamefont
  {M.}~\bibnamefont {Baenitz}}, \bibinfo {author} {\bibfnamefont
  {B.}~\bibnamefont {Schmidt}},\ and\ \bibinfo {author} {\bibfnamefont
  {P.}~\bibnamefont {Khuntia}},\ }\href
  {https://doi.org/https://doi.org/10.1016/j.physrep.2023.09.008} {\bibfield
  {journal} {\bibinfo  {journal} {Phys. Rep.}\ }\textbf {\bibinfo {volume}
  {1041}},\ \bibinfo {pages} {1} (\bibinfo {year} {2023})}\BibitemShut
  {NoStop}%
\bibitem [{\citenamefont {Wannier}(1950)}]{Wannier1950triangular}%
  \BibitemOpen
  \bibfield  {author} {\bibinfo {author} {\bibfnamefont {G.~H.}\ \bibnamefont
  {Wannier}},\ }\href {https://doi.org/10.1103/PhysRev.79.357} {\bibfield
  {journal} {\bibinfo  {journal} {Phys. Rev.}\ }\textbf {\bibinfo {volume}
  {79}},\ \bibinfo {pages} {357} (\bibinfo {year} {1950})}\BibitemShut
  {NoStop}%
\bibitem [{\citenamefont {Fazekas}\ and\ \citenamefont
  {Anderson}(1974)}]{Fazekas1974triangular}%
  \BibitemOpen
  \bibfield  {author} {\bibinfo {author} {\bibfnamefont {P.}~\bibnamefont
  {Fazekas}}\ and\ \bibinfo {author} {\bibfnamefont {P.~W.}\ \bibnamefont
  {Anderson}},\ }\href {https://doi.org/10.1080/14786439808206568} {\bibfield
  {journal} {\bibinfo  {journal} {Philos. Mag.}\ }\textbf {\bibinfo {volume}
  {30}},\ \bibinfo {pages} {423} (\bibinfo {year} {1974})}\BibitemShut
  {NoStop}%
\bibitem [{\citenamefont {Moessner}\ and\ \citenamefont
  {Sondhi}(2001)}]{Moessner2001triangular}%
  \BibitemOpen
  \bibfield  {author} {\bibinfo {author} {\bibfnamefont {R.}~\bibnamefont
  {Moessner}}\ and\ \bibinfo {author} {\bibfnamefont {S.~L.}\ \bibnamefont
  {Sondhi}},\ }\href {https://doi.org/10.1103/PhysRevB.63.224401} {\bibfield
  {journal} {\bibinfo  {journal} {Phys. Rev. B}\ }\textbf {\bibinfo {volume}
  {63}},\ \bibinfo {pages} {224401} (\bibinfo {year} {2001})}\BibitemShut
  {NoStop}%
\bibitem [{\citenamefont {Kuang}\ \emph {et~al.}(2002)\citenamefont {Kuang},
  \citenamefont {Jing}, \citenamefont {Loong}, \citenamefont {Lachowski},
  \citenamefont {Skakle},\ and\ \citenamefont {West}}]{Kuang2002growth}%
  \BibitemOpen
  \bibfield  {author} {\bibinfo {author} {\bibfnamefont {X.}~\bibnamefont
  {Kuang}}, \bibinfo {author} {\bibfnamefont {X.}~\bibnamefont {Jing}},
  \bibinfo {author} {\bibfnamefont {C.-K.}\ \bibnamefont {Loong}}, \bibinfo
  {author} {\bibfnamefont {E.~E.}\ \bibnamefont {Lachowski}}, \bibinfo {author}
  {\bibfnamefont {J.~M.~S.}\ \bibnamefont {Skakle}},\ and\ \bibinfo {author}
  {\bibfnamefont {A.~R.}\ \bibnamefont {West}},\ }\href
  {https://doi.org/10.1021/cm020374m} {\bibfield  {journal} {\bibinfo
  {journal} {Chem. Mater.}\ }\textbf {\bibinfo {volume} {14}},\ \bibinfo
  {pages} {4359} (\bibinfo {year} {2002})}\BibitemShut {NoStop}%
\bibitem [{\citenamefont {Rodriguez-Carvajal}(1990)}]{Fullprof}%
  \BibitemOpen
  \bibfield  {author} {\bibinfo {author} {\bibfnamefont {J.}~\bibnamefont
  {Rodriguez-Carvajal}},\ }in\ \href@noop {} {\emph {\bibinfo {booktitle}
  {FULLPROF: A Program for Rietveld Refinement and Pattern Matching
  Analysis}}},\ Vol.\ \bibinfo {volume} {127}\ (\bibinfo {organization}
  {Toulouse, France},\ \bibinfo {year} {1990})\BibitemShut {NoStop}%
\bibitem [{\citenamefont {Suter}\ and\ \citenamefont {Wojek}(2012)}]{MuSRfit}%
  \BibitemOpen
  \bibfield  {author} {\bibinfo {author} {\bibfnamefont {A.}~\bibnamefont
  {Suter}}\ and\ \bibinfo {author} {\bibfnamefont {B.}~\bibnamefont {Wojek}},\
  }\href {https://doi.org/https://doi.org/10.1016/j.phpro.2012.04.042}
  {\bibfield  {journal} {\bibinfo  {journal} {Phys. Procedia}\ }\textbf
  {\bibinfo {volume} {30}},\ \bibinfo {pages} {69} (\bibinfo {year}
  {2012})}\BibitemShut {NoStop}%
\bibitem [{\citenamefont {Song}\ \emph {et~al.}(2024)\citenamefont {Song},
  \citenamefont {Liu}, \citenamefont {Chen}, \citenamefont {Zhou},
  \citenamefont {Li}, \citenamefont {Tong}, \citenamefont {Wang}, \citenamefont
  {Wang}, \citenamefont {Lu}, \citenamefont {Yuan}, \citenamefont {Guo},\ and\
  \citenamefont {Tian}}]{Song2024BTNO}%
  \BibitemOpen
  \bibfield  {author} {\bibinfo {author} {\bibfnamefont {F.}~\bibnamefont
  {Song}}, \bibinfo {author} {\bibfnamefont {A.}~\bibnamefont {Liu}}, \bibinfo
  {author} {\bibfnamefont {Q.}~\bibnamefont {Chen}}, \bibinfo {author}
  {\bibfnamefont {J.}~\bibnamefont {Zhou}}, \bibinfo {author} {\bibfnamefont
  {J.}~\bibnamefont {Li}}, \bibinfo {author} {\bibfnamefont {W.}~\bibnamefont
  {Tong}}, \bibinfo {author} {\bibfnamefont {S.}~\bibnamefont {Wang}}, \bibinfo
  {author} {\bibfnamefont {Y.}~\bibnamefont {Wang}}, \bibinfo {author}
  {\bibfnamefont {H.}~\bibnamefont {Lu}}, \bibinfo {author} {\bibfnamefont
  {S.}~\bibnamefont {Yuan}}, \bibinfo {author} {\bibfnamefont {H.}~\bibnamefont
  {Guo}},\ and\ \bibinfo {author} {\bibfnamefont {Z.}~\bibnamefont {Tian}},\
  }\href {https://doi.org/10.1021/acs.inorgchem.3c04162} {\bibfield  {journal}
  {\bibinfo  {journal} {Inorg. Chem.}\ }\textbf {\bibinfo {volume} {63}},\
  \bibinfo {pages} {5831} (\bibinfo {year} {2024})}\BibitemShut {NoStop}%
\bibitem [{\citenamefont {Greedan}(2001)}]{Greedan2001exchange}%
  \BibitemOpen
  \bibfield  {author} {\bibinfo {author} {\bibfnamefont {J.~E.}\ \bibnamefont
  {Greedan}},\ }\href {DOI https://doi.org/10.1039/B003682J} {\bibfield
  {journal} {\bibinfo  {journal} {J. Mater. Chem.}\ }\textbf {\bibinfo {volume}
  {11}},\ \bibinfo {pages} {37} (\bibinfo {year} {2001})}\BibitemShut {NoStop}%
\bibitem [{\citenamefont {Bramwell}\ \emph {et~al.}(2000)\citenamefont
  {Bramwell}, \citenamefont {Field}, \citenamefont {Harris},\ and\
  \citenamefont {Parkin}}]{Bramwell2000Isingspin}%
  \BibitemOpen
  \bibfield  {author} {\bibinfo {author} {\bibfnamefont {S.~T.}\ \bibnamefont
  {Bramwell}}, \bibinfo {author} {\bibfnamefont {M.~N.}\ \bibnamefont {Field}},
  \bibinfo {author} {\bibfnamefont {M.~J.}\ \bibnamefont {Harris}},\ and\
  \bibinfo {author} {\bibfnamefont {I.~P.}\ \bibnamefont {Parkin}},\ }\href
  {https://doi.org/10.1088/0953-8984/12/4/308} {\bibfield  {journal} {\bibinfo
  {journal} {J. Phys. Condens. Matter}\ }\textbf {\bibinfo {volume} {12}},\
  \bibinfo {pages} {483} (\bibinfo {year} {2000})}\BibitemShut {NoStop}%
\bibitem [{\citenamefont {Tari}(2003)}]{tari2003specific}%
  \BibitemOpen
  \bibfield  {author} {\bibinfo {author} {\bibfnamefont {A.}~\bibnamefont
  {Tari}},\ }\href@noop {} {\emph {\bibinfo {title} {The specific heat of
  matter at low temperatures}}}\ (\bibinfo  {publisher} {Imperial College
  Press},\ \bibinfo {address} {London},\ \bibinfo {year} {2003})\BibitemShut
  {NoStop}%
\bibitem [{\citenamefont {Jiang}\ \emph {et~al.}(2022)\citenamefont {Jiang},
  \citenamefont {Yang}, \citenamefont {Gao}, \citenamefont {Wan}, \citenamefont
  {Zhu}, \citenamefont {Shiroka}, \citenamefont {Chen}, \citenamefont {Wu},
  \citenamefont {Li}, \citenamefont {Jiao}, \citenamefont {Chen}, \citenamefont
  {Bao}, \citenamefont {Tian},\ and\ \citenamefont {Shu}}]{Jiang2022BYBO}%
  \BibitemOpen
  \bibfield  {author} {\bibinfo {author} {\bibfnamefont {C.~Y.}\ \bibnamefont
  {Jiang}}, \bibinfo {author} {\bibfnamefont {Y.~X.}\ \bibnamefont {Yang}},
  \bibinfo {author} {\bibfnamefont {Y.~X.}\ \bibnamefont {Gao}}, \bibinfo
  {author} {\bibfnamefont {Z.~T.}\ \bibnamefont {Wan}}, \bibinfo {author}
  {\bibfnamefont {Z.~H.}\ \bibnamefont {Zhu}}, \bibinfo {author} {\bibfnamefont
  {T.}~\bibnamefont {Shiroka}}, \bibinfo {author} {\bibfnamefont {C.~S.}\
  \bibnamefont {Chen}}, \bibinfo {author} {\bibfnamefont {Q.}~\bibnamefont
  {Wu}}, \bibinfo {author} {\bibfnamefont {X.}~\bibnamefont {Li}}, \bibinfo
  {author} {\bibfnamefont {J.~C.}\ \bibnamefont {Jiao}}, \bibinfo {author}
  {\bibfnamefont {K.~W.}\ \bibnamefont {Chen}}, \bibinfo {author}
  {\bibfnamefont {Y.}~\bibnamefont {Bao}}, \bibinfo {author} {\bibfnamefont
  {Z.~M.}\ \bibnamefont {Tian}},\ and\ \bibinfo {author} {\bibfnamefont
  {L.}~\bibnamefont {Shu}},\ }\href
  {https://doi.org/10.1103/PhysRevB.106.014409} {\bibfield  {journal} {\bibinfo
   {journal} {Phys. Rev. B}\ }\textbf {\bibinfo {volume} {106}},\ \bibinfo
  {pages} {014409} (\bibinfo {year} {2022})}\BibitemShut {NoStop}%
\bibitem [{\citenamefont {Lhotel}\ \emph {et~al.}(2021)\citenamefont {Lhotel},
  \citenamefont {Mangin-Thro}, \citenamefont {Ressouche}, \citenamefont
  {Steffens}, \citenamefont {Bichaud}, \citenamefont {Knebel}, \citenamefont
  {Brison}, \citenamefont {Marin}, \citenamefont {Raymond},\ and\ \citenamefont
  {Zhitomirsky}}]{Lhotel2021Yb3Ga5O12}%
  \BibitemOpen
  \bibfield  {author} {\bibinfo {author} {\bibfnamefont {E.}~\bibnamefont
  {Lhotel}}, \bibinfo {author} {\bibfnamefont {L.}~\bibnamefont {Mangin-Thro}},
  \bibinfo {author} {\bibfnamefont {E.}~\bibnamefont {Ressouche}}, \bibinfo
  {author} {\bibfnamefont {P.}~\bibnamefont {Steffens}}, \bibinfo {author}
  {\bibfnamefont {E.}~\bibnamefont {Bichaud}}, \bibinfo {author} {\bibfnamefont
  {G.}~\bibnamefont {Knebel}}, \bibinfo {author} {\bibfnamefont {J.-P.}\
  \bibnamefont {Brison}}, \bibinfo {author} {\bibfnamefont {C.}~\bibnamefont
  {Marin}}, \bibinfo {author} {\bibfnamefont {S.}~\bibnamefont {Raymond}},\
  and\ \bibinfo {author} {\bibfnamefont {M.~E.}\ \bibnamefont {Zhitomirsky}},\
  }\href {https://doi.org/10.1103/PhysRevB.104.024427} {\bibfield  {journal}
  {\bibinfo  {journal} {Phys. Rev. B}\ }\textbf {\bibinfo {volume} {104}},\
  \bibinfo {pages} {024427} (\bibinfo {year} {2021})}\BibitemShut {NoStop}%
\bibitem [{\citenamefont {Khatua}\ \emph
  {et~al.}(2022{\natexlab{a}})\citenamefont {Khatua}, \citenamefont {Pregelj},
  \citenamefont {Elghandour}, \citenamefont {Jagli\ifmmode~\check{c}\else
  \v{c}\fi{}ic}, \citenamefont {Klingeler}, \citenamefont {Zorko},\ and\
  \citenamefont {Khuntia}}]{Khatua2022Ba3RB9O18}%
  \BibitemOpen
  \bibfield  {author} {\bibinfo {author} {\bibfnamefont {J.}~\bibnamefont
  {Khatua}}, \bibinfo {author} {\bibfnamefont {M.}~\bibnamefont {Pregelj}},
  \bibinfo {author} {\bibfnamefont {A.}~\bibnamefont {Elghandour}}, \bibinfo
  {author} {\bibfnamefont {Z.}~\bibnamefont {Jagli\ifmmode~\check{c}\else
  \v{c}\fi{}ic}}, \bibinfo {author} {\bibfnamefont {R.}~\bibnamefont
  {Klingeler}}, \bibinfo {author} {\bibfnamefont {A.}~\bibnamefont {Zorko}},\
  and\ \bibinfo {author} {\bibfnamefont {P.}~\bibnamefont {Khuntia}},\ }\href
  {https://doi.org/10.1103/PhysRevB.106.104408} {\bibfield  {journal} {\bibinfo
   {journal} {Phys. Rev. B}\ }\textbf {\bibinfo {volume} {106}},\ \bibinfo
  {pages} {104408} (\bibinfo {year} {2022}{\natexlab{a}})}\BibitemShut
  {NoStop}%
\bibitem [{\citenamefont {Hillier}\ \emph {et~al.}(2022)\citenamefont
  {Hillier}, \citenamefont {Blundell}, \citenamefont {McKenzie}, \citenamefont
  {Umegaki}, \citenamefont {Shu}, \citenamefont {Wright}, \citenamefont
  {Prokscha}, \citenamefont {Bert}, \citenamefont {Shimomura}, \citenamefont
  {Berlie}, \citenamefont {Alberto},\ and\ \citenamefont
  {Watanabe}}]{Hiller2002Muon}%
  \BibitemOpen
  \bibfield  {author} {\bibinfo {author} {\bibfnamefont {A.~D.}\ \bibnamefont
  {Hillier}}, \bibinfo {author} {\bibfnamefont {S.~J.}\ \bibnamefont
  {Blundell}}, \bibinfo {author} {\bibfnamefont {I.}~\bibnamefont {McKenzie}},
  \bibinfo {author} {\bibfnamefont {I.}~\bibnamefont {Umegaki}}, \bibinfo
  {author} {\bibfnamefont {L.}~\bibnamefont {Shu}}, \bibinfo {author}
  {\bibfnamefont {J.~A.}\ \bibnamefont {Wright}}, \bibinfo {author}
  {\bibfnamefont {T.}~\bibnamefont {Prokscha}}, \bibinfo {author}
  {\bibfnamefont {F.}~\bibnamefont {Bert}}, \bibinfo {author} {\bibfnamefont
  {K.}~\bibnamefont {Shimomura}}, \bibinfo {author} {\bibfnamefont
  {A.}~\bibnamefont {Berlie}}, \bibinfo {author} {\bibfnamefont
  {H.}~\bibnamefont {Alberto}},\ and\ \bibinfo {author} {\bibfnamefont
  {I.}~\bibnamefont {Watanabe}},\ }\href
  {https://doi.org/10.1038/s43586-021-00089-0} {\bibfield  {journal} {\bibinfo
  {journal} {Nat. Rev. Methods Primers}\ }\textbf {\bibinfo {volume} {2}},\
  \bibinfo {pages} {4} (\bibinfo {year} {2022})}\BibitemShut {NoStop}%
\bibitem [{\citenamefont {Zheng}\ \emph {et~al.}(2005)\citenamefont {Zheng},
  \citenamefont {Kubozono}, \citenamefont {Nishiyama}, \citenamefont
  {Higemoto}, \citenamefont {Kawae}, \citenamefont {Koda},\ and\ \citenamefont
  {Xu}}]{Zheng2005muSRoscillate}%
  \BibitemOpen
  \bibfield  {author} {\bibinfo {author} {\bibfnamefont {X.~G.}\ \bibnamefont
  {Zheng}}, \bibinfo {author} {\bibfnamefont {H.}~\bibnamefont {Kubozono}},
  \bibinfo {author} {\bibfnamefont {K.}~\bibnamefont {Nishiyama}}, \bibinfo
  {author} {\bibfnamefont {W.}~\bibnamefont {Higemoto}}, \bibinfo {author}
  {\bibfnamefont {T.}~\bibnamefont {Kawae}}, \bibinfo {author} {\bibfnamefont
  {A.}~\bibnamefont {Koda}},\ and\ \bibinfo {author} {\bibfnamefont {C.~N.}\
  \bibnamefont {Xu}},\ }\href {https://doi.org/10.1103/PhysRevLett.95.057201}
  {\bibfield  {journal} {\bibinfo  {journal} {Phys. Rev. Lett.}\ }\textbf
  {\bibinfo {volume} {95}},\ \bibinfo {pages} {057201} (\bibinfo {year}
  {2005})}\BibitemShut {NoStop}%
\bibitem [{\citenamefont {Hayano}\ \emph {et~al.}(1979)\citenamefont {Hayano},
  \citenamefont {Uemura}, \citenamefont {Imazato}, \citenamefont {Nishida},
  \citenamefont {Yamazaki},\ and\ \citenamefont {Kubo}}]{Hayano1979KT}%
  \BibitemOpen
  \bibfield  {author} {\bibinfo {author} {\bibfnamefont {R.~S.}\ \bibnamefont
  {Hayano}}, \bibinfo {author} {\bibfnamefont {Y.~J.}\ \bibnamefont {Uemura}},
  \bibinfo {author} {\bibfnamefont {J.}~\bibnamefont {Imazato}}, \bibinfo
  {author} {\bibfnamefont {N.}~\bibnamefont {Nishida}}, \bibinfo {author}
  {\bibfnamefont {T.}~\bibnamefont {Yamazaki}},\ and\ \bibinfo {author}
  {\bibfnamefont {R.}~\bibnamefont {Kubo}},\ }\href
  {https://doi.org/10.1103/PhysRevB.20.850} {\bibfield  {journal} {\bibinfo
  {journal} {Phys. Rev. B}\ }\textbf {\bibinfo {volume} {20}},\ \bibinfo
  {pages} {850} (\bibinfo {year} {1979})}\BibitemShut {NoStop}%
\bibitem [{\citenamefont {Clark}\ \emph {et~al.}(2019)\citenamefont {Clark},
  \citenamefont {Sala}, \citenamefont {Maharaj}, \citenamefont {Stone},
  \citenamefont {Knight}, \citenamefont {Telling}, \citenamefont {Wang},
  \citenamefont {Xu}, \citenamefont {Kim}, \citenamefont {Li}, \citenamefont
  {Cheong},\ and\ \citenamefont {Gaulin}}]{Clark2019TbInO3}%
  \BibitemOpen
  \bibfield  {author} {\bibinfo {author} {\bibfnamefont {L.}~\bibnamefont
  {Clark}}, \bibinfo {author} {\bibfnamefont {G.}~\bibnamefont {Sala}},
  \bibinfo {author} {\bibfnamefont {D.~D.}\ \bibnamefont {Maharaj}}, \bibinfo
  {author} {\bibfnamefont {M.~B.}\ \bibnamefont {Stone}}, \bibinfo {author}
  {\bibfnamefont {K.~S.}\ \bibnamefont {Knight}}, \bibinfo {author}
  {\bibfnamefont {M.~T.~F.}\ \bibnamefont {Telling}}, \bibinfo {author}
  {\bibfnamefont {X.}~\bibnamefont {Wang}}, \bibinfo {author} {\bibfnamefont
  {X.}~\bibnamefont {Xu}}, \bibinfo {author} {\bibfnamefont {J.}~\bibnamefont
  {Kim}}, \bibinfo {author} {\bibfnamefont {Y.}~\bibnamefont {Li}}, \bibinfo
  {author} {\bibfnamefont {S.-W.}\ \bibnamefont {Cheong}},\ and\ \bibinfo
  {author} {\bibfnamefont {B.~D.}\ \bibnamefont {Gaulin}},\ }\href
  {https://doi.org/10.1038/s41567-018-0407-2} {\bibfield  {journal} {\bibinfo
  {journal} {Nat. Phys.}\ }\textbf {\bibinfo {volume} {15}},\ \bibinfo {pages}
  {262} (\bibinfo {year} {2019})}\BibitemShut {NoStop}%
\bibitem [{\citenamefont {Uemura}\ \emph {et~al.}(1985)\citenamefont {Uemura},
  \citenamefont {Yamazaki}, \citenamefont {Harshman}, \citenamefont {Senba},\
  and\ \citenamefont {Ansaldo}}]{Uemura1985static}%
  \BibitemOpen
  \bibfield  {author} {\bibinfo {author} {\bibfnamefont {Y.~J.}\ \bibnamefont
  {Uemura}}, \bibinfo {author} {\bibfnamefont {T.}~\bibnamefont {Yamazaki}},
  \bibinfo {author} {\bibfnamefont {D.~R.}\ \bibnamefont {Harshman}}, \bibinfo
  {author} {\bibfnamefont {M.}~\bibnamefont {Senba}},\ and\ \bibinfo {author}
  {\bibfnamefont {E.~J.}\ \bibnamefont {Ansaldo}},\ }\href
  {https://doi.org/10.1103/PhysRevB.31.546} {\bibfield  {journal} {\bibinfo
  {journal} {Phys. Rev. B}\ }\textbf {\bibinfo {volume} {31}},\ \bibinfo
  {pages} {546} (\bibinfo {year} {1985})}\BibitemShut {NoStop}%
\bibitem [{\citenamefont {Yaouanc}\ and\ \citenamefont
  {De~Reotier}(2011)}]{yaouanc2011muon}%
  \BibitemOpen
  \bibfield  {author} {\bibinfo {author} {\bibfnamefont {A.}~\bibnamefont
  {Yaouanc}}\ and\ \bibinfo {author} {\bibfnamefont {P.~D.}\ \bibnamefont
  {De~Reotier}},\ }\href@noop {} {\emph {\bibinfo {title} {Muon spin rotation,
  relaxation, and resonance: applications to condensed matter}}},\ \bibinfo
  {number} {147}\ (\bibinfo  {publisher} {Oxford University Press},\ \bibinfo
  {year} {2011})\BibitemShut {NoStop}%
\bibitem [{\citenamefont {Khatua}\ \emph {et~al.}(2024)\citenamefont {Khatua},
  \citenamefont {Bhattacharya}, \citenamefont {Strydom}, \citenamefont {Zorko},
  \citenamefont {Lord}, \citenamefont {Ozarowski}, \citenamefont {Kermarrec},\
  and\ \citenamefont {Khuntia}}]{Khatua2024BTYO}%
  \BibitemOpen
  \bibfield  {author} {\bibinfo {author} {\bibfnamefont {J.}~\bibnamefont
  {Khatua}}, \bibinfo {author} {\bibfnamefont {S.}~\bibnamefont
  {Bhattacharya}}, \bibinfo {author} {\bibfnamefont {A.~M.}\ \bibnamefont
  {Strydom}}, \bibinfo {author} {\bibfnamefont {A.}~\bibnamefont {Zorko}},
  \bibinfo {author} {\bibfnamefont {J.~S.}\ \bibnamefont {Lord}}, \bibinfo
  {author} {\bibfnamefont {A.}~\bibnamefont {Ozarowski}}, \bibinfo {author}
  {\bibfnamefont {E.}~\bibnamefont {Kermarrec}},\ and\ \bibinfo {author}
  {\bibfnamefont {P.}~\bibnamefont {Khuntia}},\ }\href
  {https://doi.org/10.1103/PhysRevB.109.024427} {\bibfield  {journal} {\bibinfo
   {journal} {Phys. Rev. B}\ }\textbf {\bibinfo {volume} {109}},\ \bibinfo
  {pages} {024427} (\bibinfo {year} {2024})}\BibitemShut {NoStop}%
\bibitem [{\citenamefont {Khatua}\ \emph
  {et~al.}(2022{\natexlab{b}})\citenamefont {Khatua}, \citenamefont
  {Bhattacharya}, \citenamefont {Ding}, \citenamefont {Vrtnik}, \citenamefont
  {Strydom}, \citenamefont {Butch}, \citenamefont {Luetkens}, \citenamefont
  {Kermarrec}, \citenamefont {Rao}, \citenamefont {Zorko}, \citenamefont
  {Furukawa},\ and\ \citenamefont {Khuntia}}]{Khatua2022hyper}%
  \BibitemOpen
  \bibfield  {author} {\bibinfo {author} {\bibfnamefont {J.}~\bibnamefont
  {Khatua}}, \bibinfo {author} {\bibfnamefont {S.}~\bibnamefont
  {Bhattacharya}}, \bibinfo {author} {\bibfnamefont {Q.~P.}\ \bibnamefont
  {Ding}}, \bibinfo {author} {\bibfnamefont {S.}~\bibnamefont {Vrtnik}},
  \bibinfo {author} {\bibfnamefont {A.~M.}\ \bibnamefont {Strydom}}, \bibinfo
  {author} {\bibfnamefont {N.~P.}\ \bibnamefont {Butch}}, \bibinfo {author}
  {\bibfnamefont {H.}~\bibnamefont {Luetkens}}, \bibinfo {author}
  {\bibfnamefont {E.}~\bibnamefont {Kermarrec}}, \bibinfo {author}
  {\bibfnamefont {M.~S.~R.}\ \bibnamefont {Rao}}, \bibinfo {author}
  {\bibfnamefont {A.}~\bibnamefont {Zorko}}, \bibinfo {author} {\bibfnamefont
  {Y.}~\bibnamefont {Furukawa}},\ and\ \bibinfo {author} {\bibfnamefont
  {P.}~\bibnamefont {Khuntia}},\ }\href
  {https://doi.org/10.1103/PhysRevB.106.104404} {\bibfield  {journal} {\bibinfo
   {journal} {Phys. Rev. B}\ }\textbf {\bibinfo {volume} {106}},\ \bibinfo
  {pages} {104404} (\bibinfo {year} {2022}{\natexlab{b}})}\BibitemShut
  {NoStop}%
\bibitem [{\citenamefont {Ennis}\ \emph {et~al.}(2024)\citenamefont {Ennis},
  \citenamefont {Bag}, \citenamefont {Liu}, \citenamefont {Dissanayake},
  \citenamefont {Kolesnikov}, \citenamefont {Balents},\ and\ \citenamefont
  {Haravifard}}]{Ennis2024BEBO}%
  \BibitemOpen
  \bibfield  {author} {\bibinfo {author} {\bibfnamefont {M.}~\bibnamefont
  {Ennis}}, \bibinfo {author} {\bibfnamefont {R.}~\bibnamefont {Bag}}, \bibinfo
  {author} {\bibfnamefont {C.}~\bibnamefont {Liu}}, \bibinfo {author}
  {\bibfnamefont {S.~E.}\ \bibnamefont {Dissanayake}}, \bibinfo {author}
  {\bibfnamefont {A.~I.}\ \bibnamefont {Kolesnikov}}, \bibinfo {author}
  {\bibfnamefont {L.}~\bibnamefont {Balents}},\ and\ \bibinfo {author}
  {\bibfnamefont {S.}~\bibnamefont {Haravifard}},\ }\href
  {https://doi.org/10.1038/s42005-024-01532-w} {\bibfield  {journal} {\bibinfo
  {journal} {Communications Physics}\ }\textbf {\bibinfo {volume} {7}},\
  \bibinfo {pages} {37} (\bibinfo {year} {2024})}\BibitemShut {NoStop}%
\bibitem [{\citenamefont {Li}\ \emph {et~al.}(2016{\natexlab{b}})\citenamefont
  {Li}, \citenamefont {Wang},\ and\ \citenamefont {Chen}}]{Li2016DOdoublet}%
  \BibitemOpen
  \bibfield  {author} {\bibinfo {author} {\bibfnamefont {Y.-D.}\ \bibnamefont
  {Li}}, \bibinfo {author} {\bibfnamefont {X.}~\bibnamefont {Wang}},\ and\
  \bibinfo {author} {\bibfnamefont {G.}~\bibnamefont {Chen}},\ }\href
  {https://doi.org/10.1103/PhysRevB.94.201114} {\bibfield  {journal} {\bibinfo
  {journal} {Phys. Rev. B}\ }\textbf {\bibinfo {volume} {94}},\ \bibinfo
  {pages} {201114} (\bibinfo {year} {2016}{\natexlab{b}})}\BibitemShut
  {NoStop}%
\bibitem [{\citenamefont {Zhu}\ and\ \citenamefont {Shu}(2020)}]{Zhu2020muSR}%
  \BibitemOpen
  \bibfield  {author} {\bibinfo {author} {\bibfnamefont {Z.}~\bibnamefont
  {Zhu}}\ and\ \bibinfo {author} {\bibfnamefont {L.}~\bibnamefont {Shu}},\
  }\href {https://pip.nju.edu.cn/EN/Y2020/V40/I5/143} {\bibfield  {journal}
  {\bibinfo  {journal} {Prog. Phys.}\ }\textbf {\bibinfo {volume} {40}},\
  \bibinfo {eid} {143} (\bibinfo {year} {2020})}\BibitemShut {NoStop}%
\bibitem [{\citenamefont {Mugiraneza}\ and\ \citenamefont
  {Hallas}(2022)}]{magnetictutorial}%
  \BibitemOpen
  \bibfield  {author} {\bibinfo {author} {\bibfnamefont {S.}~\bibnamefont
  {Mugiraneza}}\ and\ \bibinfo {author} {\bibfnamefont {A.~M.}\ \bibnamefont
  {Hallas}},\ }\href {https://doi.org/10.1038/s42005-022-00853-y} {\bibfield
  {journal} {\bibinfo  {journal} {Communications Physics}\ }\textbf {\bibinfo
  {volume} {5}},\ \bibinfo {pages} {95} (\bibinfo {year} {2022})}\BibitemShut
  {NoStop}%
\bibitem [{\citenamefont {Kreitzman}\ \emph {et~al.}(1986)\citenamefont
  {Kreitzman}, \citenamefont {Brewer}, \citenamefont {Harshman}, \citenamefont
  {Keitel}, \citenamefont {Williams}, \citenamefont {Crowe},\ and\
  \citenamefont {Ansaldo}}]{kreitzman1986LCR}%
  \BibitemOpen
  \bibfield  {author} {\bibinfo {author} {\bibfnamefont {S.~R.}\ \bibnamefont
  {Kreitzman}}, \bibinfo {author} {\bibfnamefont {J.~H.}\ \bibnamefont
  {Brewer}}, \bibinfo {author} {\bibfnamefont {D.~R.}\ \bibnamefont
  {Harshman}}, \bibinfo {author} {\bibfnamefont {R.}~\bibnamefont {Keitel}},
  \bibinfo {author} {\bibfnamefont {D.~L.}\ \bibnamefont {Williams}}, \bibinfo
  {author} {\bibfnamefont {K.~M.}\ \bibnamefont {Crowe}},\ and\ \bibinfo
  {author} {\bibfnamefont {E.~J.}\ \bibnamefont {Ansaldo}},\ }\href
  {https://doi.org/10.1103/PhysRevLett.56.181} {\bibfield  {journal} {\bibinfo
  {journal} {Phys. Rev. Lett.}\ }\textbf {\bibinfo {volume} {56}},\ \bibinfo
  {pages} {181} (\bibinfo {year} {1986})}\BibitemShut {NoStop}%
\bibitem [{\citenamefont {Blundell}\ \emph {et~al.}(2022)\citenamefont
  {Blundell}, \citenamefont {Renzi}, \citenamefont {Lancaster},\ and\
  \citenamefont {Pratt}}]{Stephen2022muon}%
  \BibitemOpen
  \bibfield  {author} {\bibinfo {author} {\bibfnamefont {S.~J.}\ \bibnamefont
  {Blundell}}, \bibinfo {author} {\bibfnamefont {R.~D.}\ \bibnamefont {Renzi}},
  \bibinfo {author} {\bibfnamefont {T.}~\bibnamefont {Lancaster}},\ and\
  \bibinfo {author} {\bibfnamefont {F.~L.}\ \bibnamefont {Pratt}},\ }\href@noop
  {} {\emph {\bibinfo {title} {Muon Spectroscopy: An Introduction}}}\ (\bibinfo
   {publisher} {Oxford University Press},\ \bibinfo {year} {2022})\BibitemShut
  {NoStop}%
\bibitem [{\citenamefont {Pratt}\ \emph {et~al.}(2022)\citenamefont {Pratt},
  \citenamefont {Lang}, \citenamefont {Steinhardt}, \citenamefont
  {Haravifard},\ and\ \citenamefont {Blundell}}]{Pratt2022YZGO}%
  \BibitemOpen
  \bibfield  {author} {\bibinfo {author} {\bibfnamefont {F.~L.}\ \bibnamefont
  {Pratt}}, \bibinfo {author} {\bibfnamefont {F.}~\bibnamefont {Lang}},
  \bibinfo {author} {\bibfnamefont {W.}~\bibnamefont {Steinhardt}}, \bibinfo
  {author} {\bibfnamefont {S.}~\bibnamefont {Haravifard}},\ and\ \bibinfo
  {author} {\bibfnamefont {S.~J.}\ \bibnamefont {Blundell}},\ }\href
  {https://doi.org/10.1103/PhysRevB.106.L060401} {\bibfield  {journal}
  {\bibinfo  {journal} {Phys. Rev. B}\ }\textbf {\bibinfo {volume} {106}},\
  \bibinfo {pages} {L060401} (\bibinfo {year} {2022})}\BibitemShut {NoStop}%
\bibitem [{\citenamefont {Li}\ \emph {et~al.}(2016{\natexlab{c}})\citenamefont
  {Li}, \citenamefont {Adroja}, \citenamefont {Biswas}, \citenamefont {Baker},
  \citenamefont {Zhang}, \citenamefont {Liu}, \citenamefont {Tsirlin},
  \citenamefont {Gegenwart},\ and\ \citenamefont {Zhang}}]{Li2016musrYMGO}%
  \BibitemOpen
  \bibfield  {author} {\bibinfo {author} {\bibfnamefont {Y.}~\bibnamefont
  {Li}}, \bibinfo {author} {\bibfnamefont {D.}~\bibnamefont {Adroja}}, \bibinfo
  {author} {\bibfnamefont {P.~K.}\ \bibnamefont {Biswas}}, \bibinfo {author}
  {\bibfnamefont {P.~J.}\ \bibnamefont {Baker}}, \bibinfo {author}
  {\bibfnamefont {Q.}~\bibnamefont {Zhang}}, \bibinfo {author} {\bibfnamefont
  {J.}~\bibnamefont {Liu}}, \bibinfo {author} {\bibfnamefont {A.~A.}\
  \bibnamefont {Tsirlin}}, \bibinfo {author} {\bibfnamefont {P.}~\bibnamefont
  {Gegenwart}},\ and\ \bibinfo {author} {\bibfnamefont {Q.}~\bibnamefont
  {Zhang}},\ }\href {https://doi.org/10.1103/PhysRevLett.117.097201} {\bibfield
   {journal} {\bibinfo  {journal} {Phys. Rev. Lett.}\ }\textbf {\bibinfo
  {volume} {117}},\ \bibinfo {pages} {097201} (\bibinfo {year}
  {2016}{\natexlab{c}})}\BibitemShut {NoStop}%
\bibitem [{\citenamefont {Uemura}\ \emph {et~al.}(1994)\citenamefont {Uemura},
  \citenamefont {Keren}, \citenamefont {Kojima}, \citenamefont {Le},
  \citenamefont {Luke}, \citenamefont {Wu}, \citenamefont {Ajiro},
  \citenamefont {Asano}, \citenamefont {Kuriyama}, \citenamefont {Mekata},
  \citenamefont {Kikuchi},\ and\ \citenamefont {Kakurai}}]{Uemura1994nu}%
  \BibitemOpen
  \bibfield  {author} {\bibinfo {author} {\bibfnamefont {Y.~J.}\ \bibnamefont
  {Uemura}}, \bibinfo {author} {\bibfnamefont {A.}~\bibnamefont {Keren}},
  \bibinfo {author} {\bibfnamefont {K.}~\bibnamefont {Kojima}}, \bibinfo
  {author} {\bibfnamefont {L.~P.}\ \bibnamefont {Le}}, \bibinfo {author}
  {\bibfnamefont {G.~M.}\ \bibnamefont {Luke}}, \bibinfo {author}
  {\bibfnamefont {W.~D.}\ \bibnamefont {Wu}}, \bibinfo {author} {\bibfnamefont
  {Y.}~\bibnamefont {Ajiro}}, \bibinfo {author} {\bibfnamefont
  {T.}~\bibnamefont {Asano}}, \bibinfo {author} {\bibfnamefont
  {Y.}~\bibnamefont {Kuriyama}}, \bibinfo {author} {\bibfnamefont
  {M.}~\bibnamefont {Mekata}}, \bibinfo {author} {\bibfnamefont
  {H.}~\bibnamefont {Kikuchi}},\ and\ \bibinfo {author} {\bibfnamefont
  {K.}~\bibnamefont {Kakurai}},\ }\href
  {https://doi.org/10.1103/PhysRevLett.73.3306} {\bibfield  {journal} {\bibinfo
   {journal} {Phys. Rev. Lett.}\ }\textbf {\bibinfo {volume} {73}},\ \bibinfo
  {pages} {3306} (\bibinfo {year} {1994})}\BibitemShut {NoStop}%
\end{thebibliography}
%

\end{document}